\documentclass[aps,prb,reprint]{revtex4-1}%
\usepackage{epsfig,pslatex,latexsym,times,amssymb,amsmath,graphicx}
\usepackage{bm, float, lipsum}
\usepackage{amsmath}
\usepackage{amsfonts}
\usepackage{amssymb}
\usepackage{mathrsfs}
\usepackage{color}
\usepackage{graphicx}%
\providecommand{\U}[1]{\protect\rule{.1in}{.1in}}
\begin{document}
\author{N. J. Harmon}
\email{nicholas-harmon@uiowa.edu} 
\author{M. E. Flatt\'e}
\affiliation{Department of Physics and Astronomy and Optical Science and Technology Center, University of Iowa, Iowa City, Iowa
52242, USA}
\date{\today}
\title{Semiclassical theory of magnetoresistance in positionally-disordered organic semiconductors}
\begin{abstract}
A recently introduced percolative theory of unipolar organic magnetoresistance is generalized by treating the hyperfine interaction semiclassically for an arbitrary hopping rate. Compact analytic results for the magnetoresistance are achievable when carrier hopping occurs much more frequently than the hyperfine field precession period. 
In other regimes, the magnetoresistance can be straightforwardly evaluated numerically.
Slow and fast hopping magnetoresistance are found to be uniquely characterized by their lineshapes. We find that the threshold hopping distance is analogous a phenomenological two-site model's  branching parameter, and that the distinction between slow and fast hopping is contingent on the threshold hopping distance. 
\end{abstract}
\maketitle

\section{Introduction}

The prospect of spintronics\cite{Ziese2001,Awschalom2002,Awschalom2002SA} in organic materials\cite{Naber2007, Vardeny2010,Bergenti2011}  has generated much interest in recent years.
Spin-orbit coupling, a bane to long spin lifetimes, can often be much weaker in organic than in typical inorganic semiconductors used for electronics, due to small atomic numbers in the organic constituents. Affordable manufacturing and chemical tunability also add to the appeal of studying spins in organic systems.
In contrast to inorganic semiconductors, organic semiconductors are typically disordered and therefore their transport mobilities are much smaller. 
Despite this apparent drawback, organic semiconductors are currently used in a variety of electronic devices;\cite{Hauff2006} understanding the behavior of spins in these systems offers the possibility of adding magnetic functionality to these and future devices.
Since spin transport properties are intertwined with the electrical transport properties,\cite{Flatte2000b} features distinct from spin transport through inorganic semiconductors\cite{Awschalom2002, Awschalom2007} are expected in organics due to their very different electronic transport properties.

In parallel with these developments, researchers\cite{Kalinowski2003,Francis2004,Prigodin2006,Desai2007,Hu2007,Bloom2007,Bobbert2007,Bergeson2008,Wagemans2010, Nguyen2010} have studied a magnetic field effect in a diverse array of organic materials, the so-called organic magnetoresistance (OMAR). Ref. \onlinecite{Wagemans2010} reviews this effect. It exists in nonmagnetic materials and is characterized by magnetoresistances of 10-20\% at fields as small as 10 mT. These properties suggest applications in magnetic sensors and organic displays.\cite{Veeraraghavan2007}
OMAR has resisted explanation by  typical magnetoresistive effects such as Lorentz force deflection, wave function shrinkage, and weak (anti-)localization.\cite{Mermer2005a}
Several recent models of OMAR have been proposed. Most involve spin-dependent processes originating from hyperfine interactions and can be classified as either unipolar\cite{Bobbert2007} or bipolar\cite{Prigodin2006, Desai2007} depending on whether the OMAR mechanism relies on a single  carrier type (e-e or h-h interactions) or two  carriers types (e-h interactions).

Due to the ubiquitous disorder in organics, methods to calculate transport properties in inorganic semiconductors fail for their organic counterparts.
One common technique to evaluate transport in organic materials is percolation theory.\cite{Shklovskii1984,Baranovski2006}
Recently a model\cite{Harmon2011} has been developed of OMAR which explicitly takes into account  hopping transport for a single carrier in a disordered material using the theoretical description of percolation theory.
Here we extend the work of Ref. \onlinecite{Harmon2011} by deriving, from a semiclassical theory of the hyperfine interaction, similar results  that apply within any hopping rate regime.

The percolation model proposed in Ref. \onlinecite{Harmon2011} and further developed here reduces the complex phenomena of spin-dependent hopping to a tractable problem of $r$-percolation with an effective density of hopping-accessible sites that is modulated by magnetic fields through singlet-triplet transitions.
We focus on unipolar charge transport since several analytic results are obtainable; near the end we assess how the general features and insights of this model may shed light on bipolar magnetoresistance mechanisms. 
Inclusion of energetic disorder precludes simple results and is not treated here, although similar MR results and trends should be expected.

OMAR in unipolar transport was  studied theoretically by Bobbert \emph{et al.} in Ref. \onlinecite{Bobbert2007} and then further developed in Ref. \onlinecite{Wagemans2008} within a so-called ``two-site'' model. In the two-site model the resistance is determined by a single ``bottleneck'' pair of sites and a phenomenological branching parameter which allows carriers to circumvent the bottleneck if the bottleneck resistance becomes too large. 
An additional feature is that the two-site model requires a very large electric field to force hopping in a single direction.
Our analysis naturally accounts for bottleneck avoidance within percolation theory (no branching parameter needs to be introduced) and large electric fields are unnecessary.\cite{Arkhipov2002}
More recently, the same researchers have reexamined their two-site model numerically by solving the stochastic Liouville equation.\cite{Schellekens2011}
Our model is in qualitative agreement with the two-site model on several points as we discuss throughout this article.

Our article is organized as follows: in Section \ref{section:model} our theory is presented; we describe how transport is changed by processes that change the relative spin orientation of polaron pairs. 
Section \ref{section:percolationSection} shows how the MR is calculated from our theory.
Section \ref{section:semiclassical} identifies the hyperfine interaction as the MR mechanism and treats it semiclassically.
In Section \ref{section:fastHopping} the special case of fast hopping is examined because analytic results can be obtained.
In Section \ref{section:general}, MR is investigated for arbitrary hopping rates.
Section \ref{section:bipolar} examines how our theory may pertain to bipolar organic devices.

\section{Model}\label{section:model}

A spatially disordered organic system can be modeled as a network of random resistors as described by Miller and Abrahams for doped inorganic semiconductors.\cite{Miller1960} The inter-site hopping resistance between two sites, $i$ and $j$, is given by $R_{ij} = R_0 e^{2 r_{ij}/\ell}$ where $r_{ij}$ is the inter-site separation and $\ell$ is the localization length of a carrier at each site. 
For simplicity, the localization length is taken to be uniform throughout the site array.
Percolation theory offers a method to calculate the bulk resistance in such a random resistor network.\cite{Ambegaokar1971, Pollak1972, Shklovskii1984} 
A critical resistance (distance) $R_c$ ($r_c$), which is the smallest resistance (or equivalently the smallest separation) that still allows for an infinitely large network of bonds, sets the bulk resistance. 
This percolation length is governed by a bonding criterion: 
\begin{equation}\label{eq:bonding1}
B_c = 4 \pi \int_0^{r_c}r^2 N dr,
\end{equation}
where $N$ is the density of sites and $B_c \approx 2.7$ in three dimensions is a number that determines the average number of bonds each site in the percolating network must connect with.\cite{Shklovskii1984}
This ``$r$-percolative'' transport model is valid when the inter-site separation is large and temperatures are high, and has been observed in organic semiconductors.\cite{Gill1974, Rubel2004}
We do not treat here smaller inter-site separations and lower temperatures where energy disorder is vital to understanding transport.
In principle, the theory here can be generalized to treat energy disorder.
 
\begin{figure}[ptbh]
 \begin{centering}
        \includegraphics[scale = 0.32,trim = 80 140 80 200, angle = -0,clip]{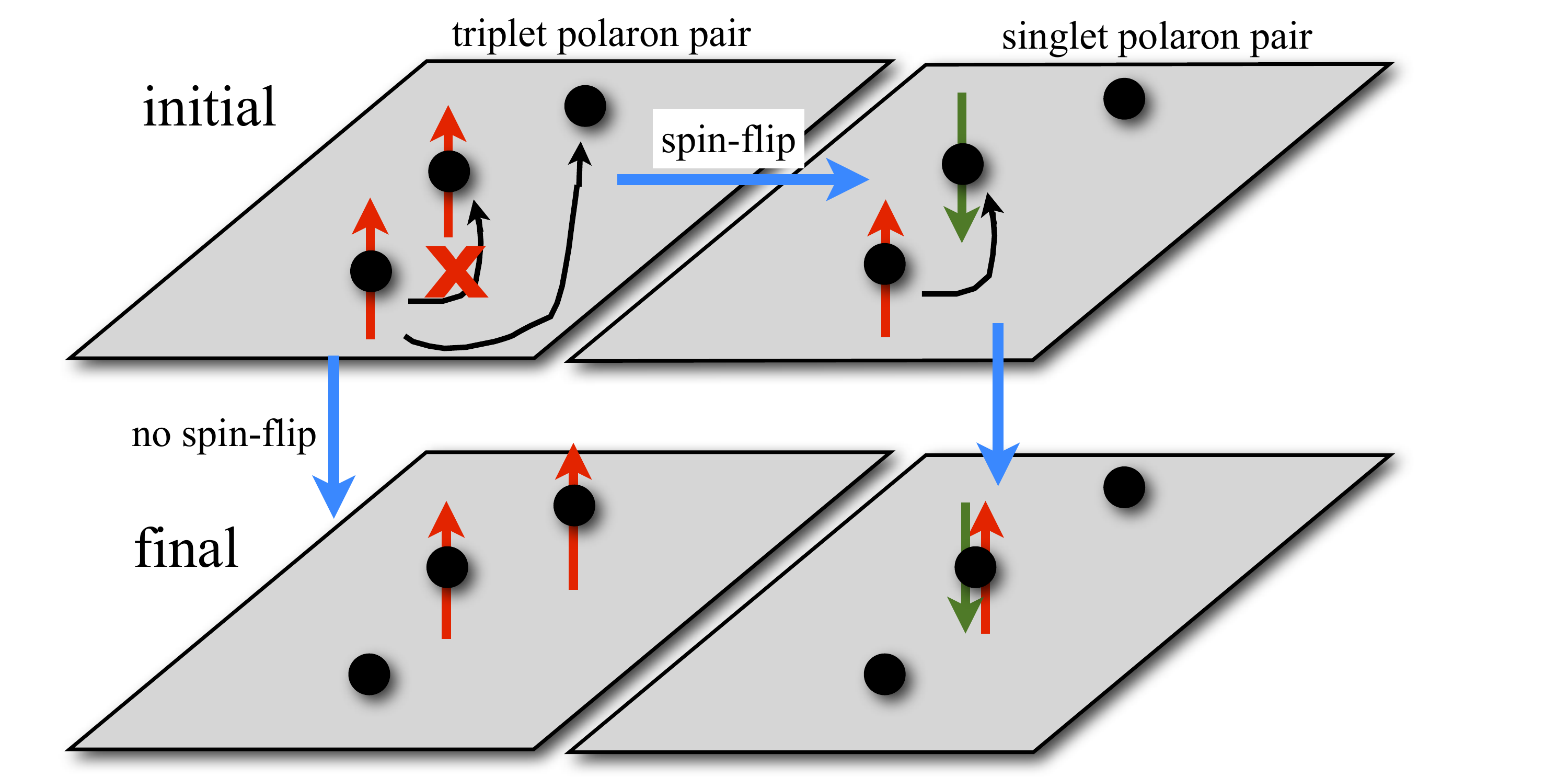}
        \caption[]
{Spin blocking in transport. 
Top: the initial situation for hopping a triplet polaron pair (left) and a singlet polaron pair (right).  Bottom: on the left  the spin-blocked carrier has made the more difficult further hop to an unoccupied site. On the right, the carrier successfully hops to the occupied nearest neighbor since the pair has a singlet spinor. The carrier concentration is dilute enough such that sites near the polaron pair are unoccupied.}\label{fig:cartoon1}
        \end{centering}
\end{figure}

Figure \ref{fig:cartoon1} displays how the Pauli exclusion principle affects spin transport in hopping conduction; double occupation at a single site is forbidden if their spins form a triplet state (T), but permissible if in the singlet state (S).\cite{Bussac1993, Bobbert2007} 
Following many of the earlier works on OMAR,\cite{Wagemans2010} we use an alternate terminology from Ref. \onlinecite{Harmon2011}, and  describe carriers in conjunction with their localizing sites as polaron quasiparticles.
An arbitrary spin localized at a site (a polaron) is unable to hop to another polaron's site if the polaron pair's (PP) spin state is T,  but can hop to site if the spin pair forms an S state just as it would to an unoccupied site, as schematically shown in Figure \ref{fig:cartoon1}. 
The respective concentrations of these three types of sites are $N_{T}$, $N_{S}$, and $N_{0}$. 
Furthermore, spin statistics dictates that $N_S = \frac{1}{4} N_1$ and $N_T = \frac{3}{4} N_1$ with $N_1$ being the concentration of singly occupied sites. 
The concentration of carriers is small enough such that a polaron encountering a bipolaron is extremely unlikely;
additionally, PPs are sufficiently separated by unoccupied sites. 

In physical systems, double occupation of a site costs a Coulomb interaction energy $U$.
Within the $r$-percolation picture, $U > 0$ inhibits double occupation and reduces any spin-dependent magnetoresistive effects.
However, in more realistic systems when energetic site disorder is larger or on the order of $U$, the effect of $U$ is not as straightforward. 
In such a case, hopping does not occur between sites with identical energies but between sites with energy difference $U$.
Surprisingly, theory for MR in which there is also energy disorder entails larger MR effects for positive non-zero $U$ than for $U = 0$.\cite{Osaka1979} 
Since we only consider positional disorder we assume $U=0$ to avoid an unphysical inhibition of double occupation. 

\begin{figure}[ptbh]
 \begin{centering}
        \includegraphics[scale = 0.25,trim = 0 15 40 10, angle = -0,clip]{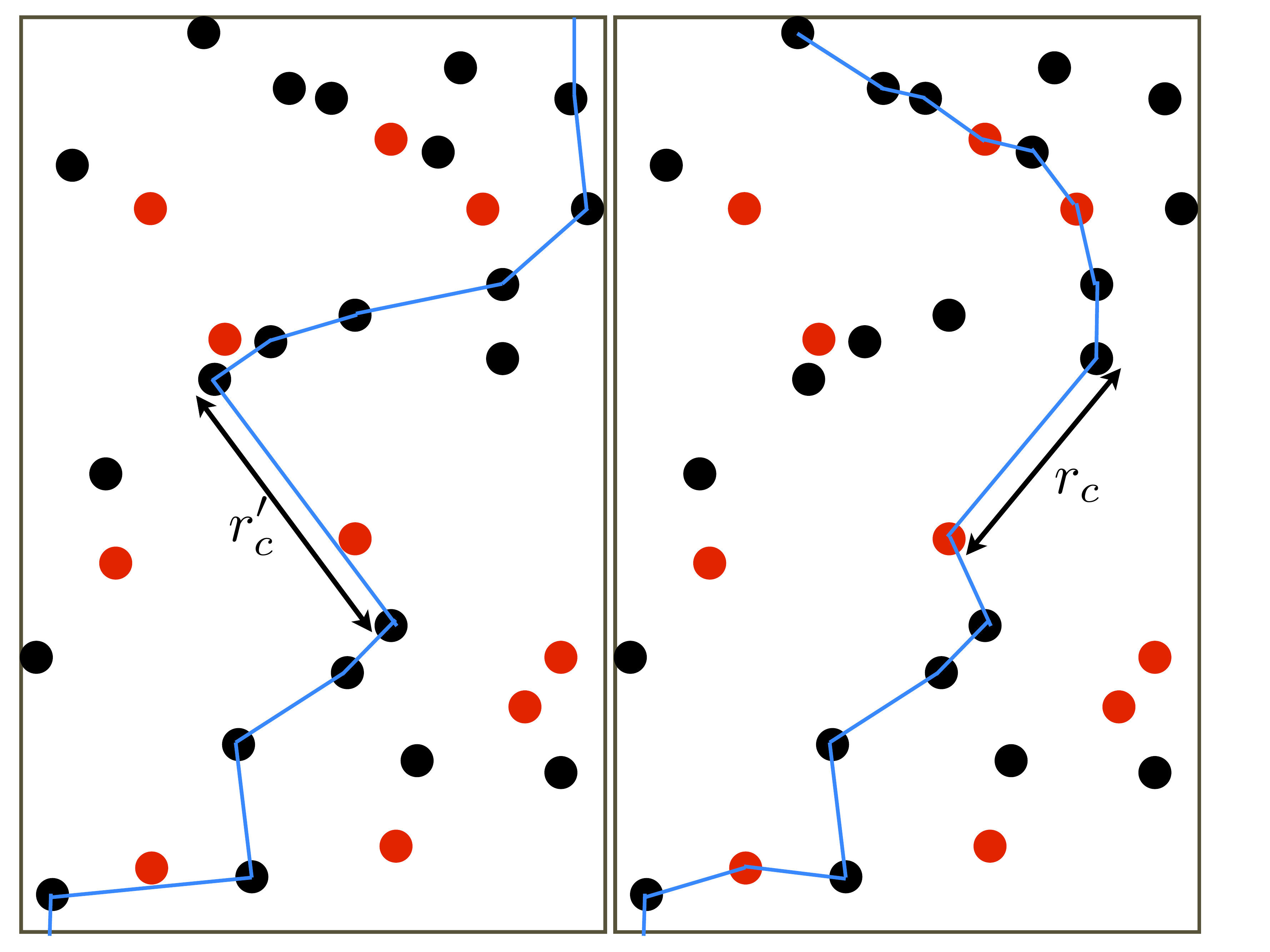}
        \caption[]
{Spin-blocking in percolative transport. 
A carrier spin starts at the bottom and begins nearest neighbor hops to the top (along solid blue line). 
Hopping-accessible sites with density $N_{eff}'$ are denoted by black solid circles. 
Left: Red sites are completely inaccessible due to spin-blocking when no spin transitions take place (as described in the main text). 
The resistance is determined by the critical hop of $r_c$. 
Right: when spin transitions occur, some red sites become accessible making the total density of hopping-accessible sites $N_{eff}$. 
The result of spin transitions is that the average inter-site separation and $r_c$ decreases which modifies the hopping path.}\label{fig:cartoon2}
        \end{centering}
\end{figure}

Since bipolaron formation is forbidden in the T states, the concentration of sites is effectively reduced to $N_{eff}' = N  - N_T$ since only these sites are accessible to a hopping polaron. In the absence of T-S transitions (or spin flips in the language of Ref. \onlinecite{Harmon2011}), we would then rewrite the bonding criterion of Eq. (\ref{eq:bonding1}) with $N\rightarrow N_{eff}'$.
A PP's hopping dynamics is thus strongly dependent on T-S transitions since bipolaron formation is completely blocked for T-states and allowable for S-states.
The effective reduction of site density entails that in general longer hops need to be achieved and an overall increase in resistance is expected as shown in Figure \ref{fig:cartoon2} (a).
If the total concentration of singly occupied sites is fixed at $N_1$, the probability of a successful hop to an occupied site (hopping to an occupied site happens with probability $N_1/N$) is 1/4, independent of spin effects. 
So, one-quarter the time the hop will be successful and the total density of sites for which successful hops take place is $N_0 + N_{S}$. 
So as before the density of unrestricted hopping sites is $N_{eff}'$. 
We must now account for the situation that occurs the other three-quarters time in which the hopping attempt to a singly occupied site is foiled due to occupation by a T-forming spin, which occurs at $N_T$ sites. 

We introduce the possibility that the spin-blocked path can be opened by a process that alters the PP's spin state; namely transitions from T to S. 
The probability for the blockade to be lifted by the time the next hopping attempt takes place, $\tau_h$, is $p_{T\rightarrow S}$. 
We thus modify the effective density of T sites  to be $[1-p_{T\rightarrow S}] N_T$.
The bonding criterion becomes
\begin{equation}\label{eq:bondingCriterion1}
B_c = 4 \pi \int_0^{r_c}r^2 N_{eff} dr,
\end{equation}
where further $r$ dependence lies in 
\begin{equation}\label{eq:effectiveDensity}
N_{eff}= N  - N_T + p_{T \rightarrow S} N_T
\end{equation}
through the hopping time $\tau_h$. 
Our model displays an interplay for a PP of two events: waiting for the transition to S to hop to the nearest site versus disassociation by hopping to a  site further away.

\section{Calculating the magnetoresistance for spatially disordered organic systems}\label{section:percolationSection}

Eq. (\ref{eq:bondingCriterion1}) is the starting point for deriving the magnetoresistance.
As discussed, the effective density to be used is $N_{eff}$, which yields
\begin{equation}\label{eq:bondingCriterion}
B_c =  \frac{4}{3} \pi \ell^3 y_c^3 (N - N_T) + 4 \pi \ell^3 N_T \int_0^{y_c}  y^2 p_{T \rightarrow S} dy,
\end{equation}
where $y_c = r_c/\ell$ is the dimensionless critical length which dictates the threshold resistance $R_c = R_0 e^{2 y_c}$; $\tau_h =  v_0^{-1} e^{2 y}$ is the hopping time. A quantity $y_{c_0} = (3 B_c/4 \pi \ell^3 N)^{1/3}$ is defined as the critical inter-site spacing in the absence of all spin effects.
 In general, $y_c$ cannot be isolated in Eq. (\ref{eq:bondingCriterion}) and the resultant MR can only be obtained  numerically unless the system is in the dilute carrier regime ($N_1 \ll N$) (which is what is assumed throughout this article).

To solve for the MR we first need to find the critical length $y_c$:
\begin{equation}
 \frac{4}{3} \pi \ell^3 y_c^3 (N - N_T)   =  B_c- 4 \pi \ell^3 N_T \int_0^{y_c}  y^2 p_{T \rightarrow S} dy,
\end{equation}
\begin{eqnarray}
 y_c^3   &=&  \frac{3 B_c}{4 \pi \ell^3(N- N_T)}- \frac{3 N_T}{N-N_T} \int_0^{y_c}  y^2 p_{T \rightarrow S} dy{}\nonumber\\
{}&=&  y_{c_1}^3 - \frac{3 N_T}{N-N_T} \int_0^{y_c}  y^2 p_{T \rightarrow S} dy.
\end{eqnarray}
where $y_{c_1} = y_{c_0} (1-N_T/N)^{-1/3}$ is the renormalized critical inter-site spacing.
$y_c$ is near $y_{c_1}$ since the singly occupied site concentration is small. 
So on the right hand side we can approximate $y_c \sim y_{c_1}$ and $N-N_T \sim N$ leaving us with
\begin{eqnarray}
 y_c  &=& y_{c_1}\Big(1 - \frac{1}{y_{c_1}^3}\frac{3 N_T}{N} \int_0^{y_{c_1}}  y^2 p_{T \rightarrow S} dy\Big)^{1/3}{}\nonumber\\
 {}&\approx&
y_{c_1} - \frac{1}{y_{c_1}^2}\frac{N_T}{N} \int_0^{y_{c_1}}  y^2 p_{T \rightarrow S} dy.
\end{eqnarray}
We see that by incorporating T$\rightarrow$S transitions, the critical length decreases from the length where triplet sites are completely excluded.
$R_c$ is then 
\begin{eqnarray}
R_c &=& R_0 e^{2 y_c} = R_0 e^{2 y_{c_1}}e^{-\frac{2}{y_{c_1}^2}\frac{N_T}{N} \int_0^{y_{c_1}}  y^2 p_{T \rightarrow S} dy} {}\nonumber\\
{}&\approx&
R_0 e^{2 y_{c_1}} (1  -  \frac{2}{y_{c_1}^2}\frac{ N_T}{N} \int_0^{y_{c_1}}  y^2 p_{T \rightarrow S} dy);
\end{eqnarray}
the MR is
\begin{eqnarray}
\textrm{MR} &\equiv& \frac{R_c(B) - R_c(0)}{R_c(0)} {}\nonumber\\
{}&&\approx 
 \frac{2}{y_{c_1}^2}\frac{N_T}{N} \int_0^{y_{c_1}}  y^2 [p_{T \rightarrow S}(0)  - p_{T \rightarrow S}(B) ]dy
\end{eqnarray}
which was first found in Ref. \onlinecite{Harmon2011}.
The problem now reduces to finding the probabilities for singlets at the next hop that were initiated in the singlet state.
However not all hops happen exactly at $\tau_h$ but over an exponential distribution of hopping times;\cite{Schulten1978, Wagemans2011a} to account for this, we write
$p_{T\rightarrow S}(B) = \frac{1}{\tau_h} \int_0^{\infty} \rho_{T\rightarrow S}(B, t) e^{-t/\tau_h} dt$
where $\rho_{T\rightarrow S}(B, t)$ is the density matrix element signifying the occupation probability of the singlet state. 
This quantity $\rho_{T\rightarrow S}$ can be related to an easier to calculate quantity $\rho_{S\rightarrow S}$ which is the population fraction in the singlet state that were initially in the singlet state.  
Their relation is $\frac{1}{3} (1 - \rho_{S\rightarrow S}(t))$.\cite{Werner1977}
In summary, we find
\begin{eqnarray}\label{eq:fullMR}
\textrm{MR}  &=& \frac{2}{3} \frac{1}{y_{c_1}^2} \eta \times{}\\
{}&&\int_0^{y_{c_1}}  y^2  \int_0^{\infty} \frac{1}{\tau_h} [\rho_{S\rightarrow S}(B)  - \rho_{S\rightarrow S}(0) ] e^{-t/\tau_h} \operatorname dt \operatorname dy,\nonumber
\end{eqnarray}
where $\eta = N_T/N$.
Due to its frequent use, $\rho_{S\rightarrow S}$ will be now denoted by the simpler $\rho_{S}$.
To calculate the magnetoresistance, the singlet population remaining after a time $t$ must be determined given various spin interactions.
The interactions considered in the follow sections are the Zeeman and hyperfine interactions. 
Spin-spin interactions (exchange and dipolar) are considered elsewhere.\cite{Harmon2012}

\section{Semiclassical model with nuclear moment at both sites}\label{section:semiclassical}

\begin{figure}[ptbh]
 \begin{centering}
        \includegraphics[scale = 0.5,trim = 400 290 80 170, angle = -0,clip]{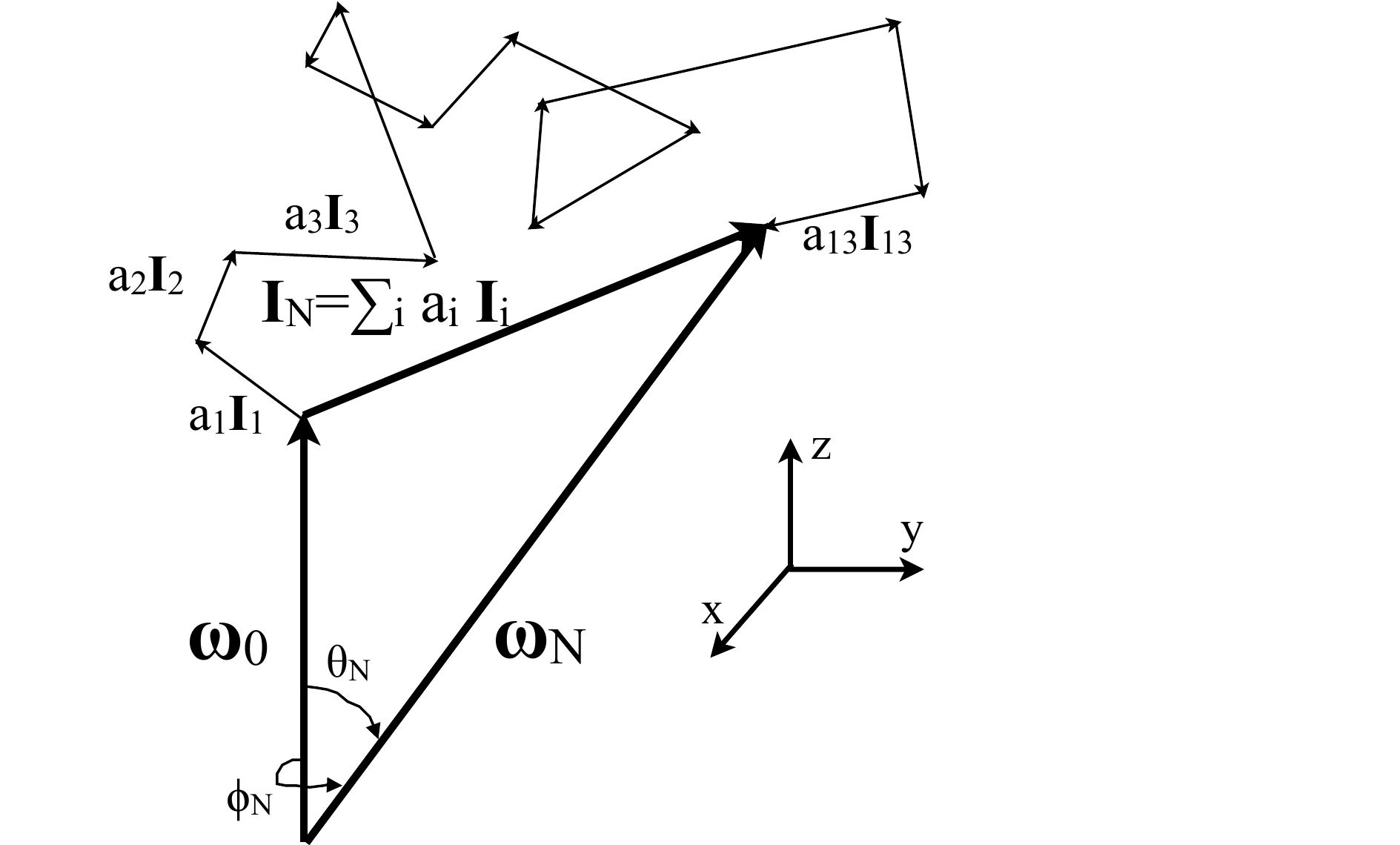}
        \caption[]
{Semiclassical description of total field. Each site is composed of some number of spin-moment-carrying nuclei which combine with the external field $\bm{\omega}_0$ to create $\bm{\omega}_N$. Each site has a different total field picked from the distribution of Eq. (\ref{eq:hfDist}).}\label{fig:hyperfineDiagram}
        \end{centering}
\end{figure}

Now the mechanisms by which T-S transitions take place are described. The physical picture is that of a PP composed of two spin-$\frac{1}{2}$ carriers located at two sites. 
While the spins are localized they evolve coherently under the influence of identical applied fields and different hyperfine fields. 
The semiclassical approximation entails that hyperfine or nuclear spins are accounted for by classical magnetic fields.  
The hyperfine field at a site is composed of many different nuclei as depicted in Figure \ref{fig:hyperfineDiagram}; however since the nuclear spin precession is so slow, the total nuclear field is assumed constant throughout the polaron's time of residence at that particular site.
Since there is no nuclear spin order the orientation and magnitude of the total hyperfine field varies from site to site. 
When one of the polarons hops, its coherent spin evolution ceases as the polaron now feels a new local magnetic field (if disassociating out of the PP by hopping to an unoccupied site) or exists as a bipolaron (if hopping to the other polaron's site) and is necessarily in the singlet state (a large exchange interaction prevents further spin evolution due to the proximity of the two polarons). If the hopping is fast, the PP spin has had little time to evolve. If the hopping is slow, each spin of the PP can be thought of as having precessed many times around its local magnetic field.

The mathematical details of the semiclassical approximation now follow. 
The Larmor frequency of a polaron spin localized at a site due to the nuclear conglomerate spin is the \emph{constant} classical vector
\begin{equation}
\bm{I}_{N} = \sum_j a_j \bm{I}_j
\end{equation}
where $a_j$ is the hyperfine coupling constant in angular frequency units between the electron and the $j$-th nucleus.
$\bm{I}_{N}$ is constructed from many nuclei each with vector length $a_j \sqrt{I_j(I_j+1)}$ pointing in a random direction. $I_j$ is the spin quantum number. 
The probability distribution for finding a specific site among an ensemble of such sites with its total end-to-end vector between $\bm{I}_{N}+d\bm{I}_{N}$ is \cite{Flory1969,Schulten1978}
\begin{equation}\label{eq:flory}
W(\bm{I}_{N}) = ({4 \pi a_{N}^2})^{-3/2} \exp\left[{-(I^2_{N}/4a^2_{N})}\right]
\end{equation}
where
\begin{equation}
a^2_{N} = \frac{1}{6} \sum_j a_j^2 I_j(I_j+1) .
\end{equation}
The \emph{effective} hyperfine coupling due to all the nuclei at a site is $a_{eff} = 2 \sqrt{2} a_N$ (although other conventions do exist\cite{Rodgers2007}).
$a_{eff}$ could be different for different types of molecular sites in which case it would have to be labeled by a site index (we neglect this effect here).
The effective hyperfine magnetic \emph{field} is $B_{eff} = a_{eff}/\gamma_e$ with the electronic gyromagnetic ratio being $\gamma_e = 0.176$ ns$^{-1}$ mT$^{-1}$. 
The effective magnetic field  is on the order of 1 mT or $a_{eff} \sim 0.176$ ns$^{-1}$ for many organic systems demonstrating OMAR.

The total precession rate seen by a carrier at a single site is then 
\begin{equation}
\bm{\omega}_{N} = \bm{I}_{N}  + \bm{\omega}_0,
\end{equation}
where $\bm{\omega}_0 = \gamma_e B\hat{z}$ is the applied field.
So the PP Hamiltonian is
\begin{equation}\label{eq:totalHam}
\mathscr{H} 
 = 
 \bm{\omega}_{N_1} \cdot \bm{S}_1 +\bm{\omega}_{N_2} \cdot \bm{S}_2,
\end{equation}
where site indices have been incorporated.
To account for the ensemble of PPs the singlet probability is averaged over the distribution $W$.
In Section \ref{section:general}, for computational reasons, the scheme of Ref. \onlinecite{Knapp1979} is followed by integrating over the \emph{total} field ($\omega_N$) as opposed to the hyperfine field.
To aid this the probability distribution is rewritten as
\begin{equation}\label{eq:hfDist}
W(\bm{\omega}_{N}) = \Big( \frac{1}{4 \pi a_{N}^2} \Big)^{3/2} \exp\Big(-\frac{1}{4}\frac{\omega_0^2 + \omega_N^2 - 2 \omega_0 \omega_N \cos\theta_N}{a_{N}^2}\Big)
\end{equation}
with differential volume element $\sin\theta_N \operatorname d\theta_N \omega_N^2 \operatorname d\omega_N \operatorname d\phi_N$.
To find the MR, determinations of the singlet density matrix elements must be made.

\section{Fast hopping}\label{section:fastHopping}

Before examining a theory applicable to any hopping time and field strength, we first explore the fast hopping regime for which simple and tractable analytic results can be obtained. The results derived herein allow us to confirm the validity of the spin relaxation model  proposed in Ref. \onlinecite{Harmon2011} over the entire range of magnetic fields. 

The dynamics for small $a_{N} t$  can be solved analytically for  arbitrary $\omega_0 t$ using perturbation theory in the interaction representation.\cite{Haberkorn1977} 
First Eq. (\ref{eq:totalHam}) is rewritten as $\mathscr{H} = \mathscr{H}_{hf} + \mathscr{H}_Z$ where
\begin{equation}\label{eq:hfHam}
\mathscr{H}_{hf}
 = 
\bm{I}_{N_1} \cdot \bm{S}_1 +  \bm{I}_{N_2} \cdot \bm{S}_1,
\end{equation}
and
\begin{equation}\label{eq:zHam}
\mathscr{H}_{Z}
 = 
\bm{\omega}_{0} \cdot \bm{S}_1+ \bm{\omega}_{0} \cdot \bm{S}_2
\end{equation}
are the hyperfine and Zeeman Hamiltonians, respectively.
In the interaction representation, the following operators are defined:
\begin{equation}
\mathscr{H}_{hf}^{*}(t) = e^{i \mathscr{H}_Z t/\hbar} \mathscr{H}_{hf} e^ {-i \mathscr{H}_Z t/\hbar},
\end{equation}
\begin{equation}
\rho^{*}(t) = e^{i \mathscr{H}_Z t/\hbar} \rho(t) e^ {-i \mathscr{H}_Z t/\hbar};
\end{equation}
initially it can be shown that $\rho^{*}(0) = \rho(0)$.
Time dependent perturbation theory for the density matrix to second order gives\cite{Slichter1963}
\begin{eqnarray}
\rho^{*}(t) &=& \rho(0) + \frac{i}{\hbar} \int_0^t [\rho(0), \mathscr{H}_{hf}^{*}(t')]dt' -{}\nonumber\\
{}&&\frac{1}{\hbar^2} \int_0^t \int_0^{t'}[[\rho(0), \mathscr{H}_{hf}^{*}(t')], \mathscr{H}_{hf}^{*}(t'')]dt'' dt'.
\end{eqnarray}
The new hyperfine Hamiltonian is 
\begin{equation}
\mathscr{H}_{hf}^{*}(t)  =  U(\mathscr{H}_{{hf}_1}+\mathscr{H}_{{hf}_2})U^{\dagger},
\end{equation}
 with 
\begin{equation}
U = e^{i \mathscr{H}_Z t/\hbar} = [\cos\frac{\omega_0 t}{2} + 2 i S_1^z \sin\frac{\omega_0t}{2}] [\cos\frac{\omega_0 t}{2} + 2 i S_2^z \sin\frac{\omega_0 t}{2}],
\end{equation}
because $\bm{S}_1$ commutes with $\bm{S}_2$.
In our singlet/triplet basis we write $\rho(0) = P_S$, which is the singlet projection operator written in Appendix \ref{app:spinOperators}.
The singlet part of the density matrix $\langle S | \rho^{*}(t)|S \rangle$ is  $\rho_S(t)$.
Initialization in the singlet state requires that $\rho_S(0) = 1$.
The first order correction vanishes.
After averaging over the hyperfine fields for the ensemble of two carriers, including the second order term yields
\begin{equation}\label{eq:arbField}
\rho_S = 1 - \frac{1}{16} (a_{eff,1}^2+a_{eff,2}^2) t^2 \bigg[1+2 \frac{\sin^2 \omega_0 t/2}{(\omega_0 t/2)^2}\bigg]
\end{equation}
in agreement with the quantum mechanical calculation.\cite{Haberkorn1977, Salikhov1984}
\begin{figure}[ptbh]
 \begin{centering}
        \includegraphics[scale = 0.65,trim = 0 0 0 0, angle = -0,clip]{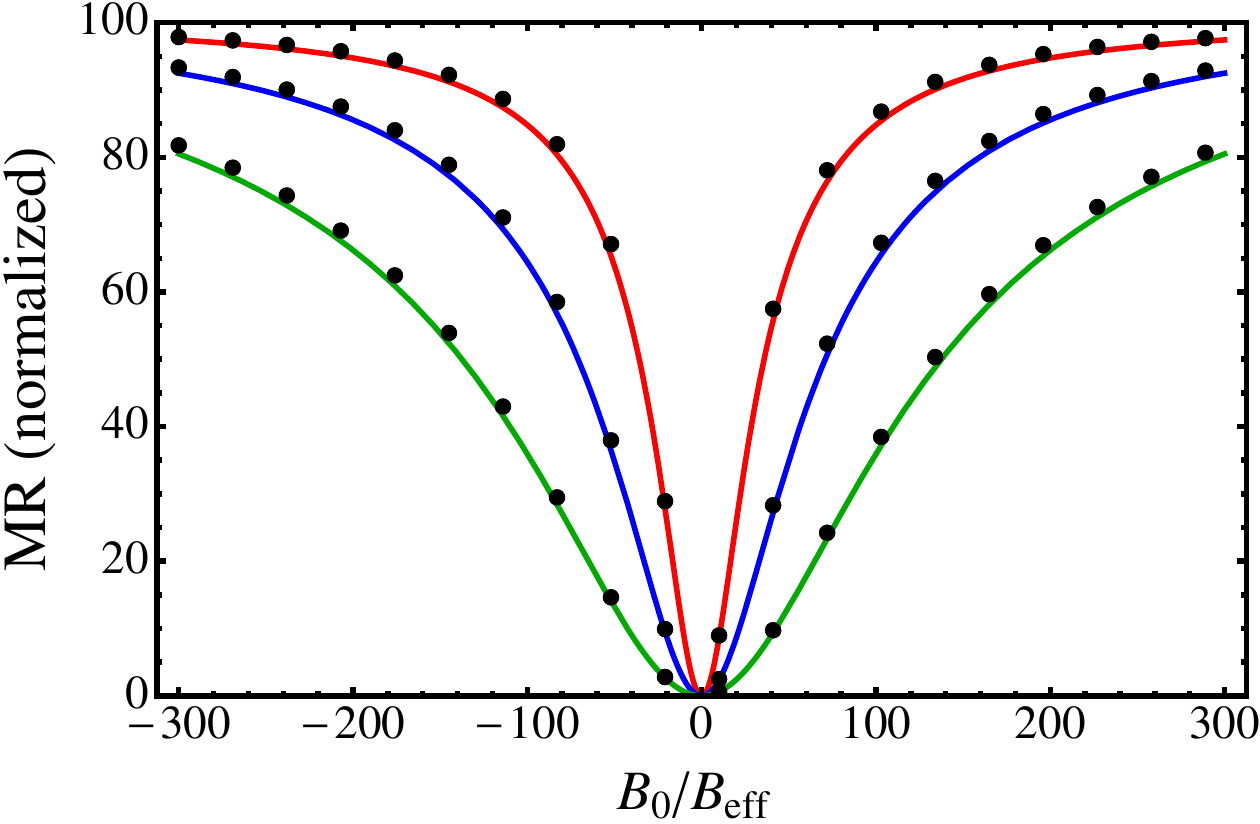}
        \caption[]
{Fast hopping normalized magnetoresistance for $y_c = 5$. Red: $r_0 = 0.5 \times 10^6$; blue: $r_0 = 1 \times 10^6$; green: $r_0 = 2 \times 10^6$. Accompanying points are calculations from general theory of Section \ref{section:general}. Not shown in plot  is that the saturated MR strongly reduces with  an increased hopping rate [which can be seen from Eq. (\ref{eq:osaka})].}\label{fig:fastHopping2}
        \end{centering}
\end{figure}

To find the MR, we substitute the singlet probability, Eq. (\ref{eq:arbField}), into Eq. (\ref{eq:fullMR}). 
After integrating over the exponential distribution of hopping times,
\begin{equation}\label{eq:MR1}
MR =\frac{1}{6} \eta
\frac{1}{ y_{c_1}^2} \int_0^{y_{c_1}}  \frac{(a_{eff,1}^2 + a_{eff,2}^2) \omega_0 ^2 \tau_h^2}{\omega_0 ^2 + 1/\tau_h^2} y^2 dy.
\end{equation}
This results agrees with the spin relaxation result\cite{Harmon2011} to within a numerical factor.
The agreement validates the interpretation given by the spin relaxation model; the intersystem crossing between singlets and triplets can be described by spin relaxation due to the rapidly varying hyperfine interaction (due to the fast hopping) in the motional narrowing regime. The spin relaxation rate is taken to be
\begin{equation}\label{eq:spinRelaxation}
\tau_s^{-1} = a_{eff}^2 \tau_h^{-1}/(\omega_0^2+\tau_h^{-2}).
\end{equation} 
The probability for the spin-flip is $1-\exp(-\tau_h/\tau_s) \approx \tau_h/\tau_s$.

The integral in Eq. (\ref{eq:MR1}) can be computed when the hopping rate has an exponential dependence on the hopping distance $\tau_h^{-1} = v_0 \exp(-2 y)$.
The result is cumbersome but can be considerably simplified under the usual assumption that $y_c \gg 1$ to
\begin{equation}\label{eq:osaka}
\textrm{MR} =
\frac{1}{24} \eta (a_{eff,1}^2 + a_{eff,2}^2) \tau^2_c\big[1 - \frac{1}{\omega_0 ^2 \tau^2_c}\ln(1+\omega_0^2 \tau^2_c) \big],
\end{equation}
where $\tau_c^{-1} = v_0 \exp{(-2 y_c)}$.
At large fields, the MR saturates at MR$_{sat} = \frac{1}{24}  \eta(a_{eff,1}^2 + a_{eff,2}^2) \tau^2_c$.
Figure \ref{fig:fastHopping2} shows several instances of the fast hopping MR. 
The following general features should also be noted. First, the saturated MR is dependent on the hyperfine field and the hopping rate;
the MR decreases as hopping times shorten, as the hyperfine fields have less time to mix the triplet to the singlet state. Second, the MR lineshape is \emph{independent} of the hyperfine field and solely dependent on the hopping rate.
The width increases with increases in the hopping rate; larger fields are required to suppress the hyperfine fields as evident from the motional narrowing spin relaxation formula Eq. (\ref{eq:spinRelaxation}). Finally, the derived MR is always positive; the applied field suppresses T-S mixing.
These features were pointed out in Ref. \onlinecite{Harmon2011} and were also confirmed by numerical simulations solving the stochastic Liouville equation.\cite{Schellekens2011}

\section{Arbitrary Hopping}\label{section:general}

The results of the previous section are valid only for short hopping times.
A different approach must be utilized to evaluate the MR for long hopping times.
Initially the density matrix is $\rho(0) = |S \rangle \langle S |$.
At some later time, under the evolution of the Hamiltonian, $\mathscr{H}$, 
\begin{eqnarray}\label{eq:densityMatrix}
\rho(t) &=& \exp(-i \mathscr{H} t/\hbar) \rho(0) \exp(i \mathscr{H} t/\hbar){}\nonumber\\
 {}&=&  \exp(-i \mathscr{H} t/\hbar) |S \rangle \langle S |  \exp(i \mathscr{H} t/\hbar).
\end{eqnarray}
We are interested in the S portion of the density matrix $\rho_{S} =\langle S | \rho(t) | S\rangle$
which then becomes
\begin{eqnarray}\label{eq:semiclassical}
\rho_{S} &=& \langle S| \exp(-i \mathscr{H} t/\hbar) |S \rangle \langle S|  \exp(i \mathscr{H} t/\hbar) | S \rangle {} \\ \nonumber
{} &=& |\langle S| \exp(-i \mathscr{H} t/\hbar) |S \rangle|^2.
\end{eqnarray}
An average over the nuclear field distribution is taken to account for the ensemble of carriers at sites with differing nuclear configurations.
Under certain restrictions for the nuclei, the problem can also be solved quantum mechanically.\cite{Salikhov1984,Timmel1998}
However when many nuclei are present at each site, which is the case in disordered organic semiconductors, the quantum mechanical calculation is best tackled numerically. 
As expected, it has been shown that the validity of the semiclassical approximation improves with the number of nuclei.\cite{Rodgers2007} 

Eq. (\ref{eq:semiclassical}) is solved by noting that our Hamiltonian, Eq. (\ref{eq:totalHam}), is Zeeman-like so we can use:
\begin{equation}\label{eq:spinRotation}
e^{-i c \hat{n} \cdot \bold{S}} = \cos\frac{c}{2} - 2 i \hat{n} \cdot \bold{S} \sin\frac{c}{2}.
\end{equation}
In Eq. (\ref{eq:spinRotation}), we have the \emph{total} field at a site ($\bm{\omega}_{N}$) unit vector
\begin{equation}
n_x  = \sin\theta_{N} \cos\phi_{N}, ~ n_y = \sin\theta_{N} \sin\phi_{N}, ~ n_z =  \cos\theta_{N},
\end{equation}
with angles defined in Figure \ref{fig:hyperfineDiagram}.
With some labor, it can be shown that\cite{Salikhov1984}
\begin{equation}\label{eq:salikhov}
\rho_S = F_1(1) F_1(2)+2 F_2(1) F_2(2)+F_3(1) F_3(2)+ \frac{1}{2}F_4(1) F_4(2),
\end{equation}
where
\begin{equation}
F_1(i)  = 1 -\langle \sin^2 (\frac{\omega_N t}{2})\rangle_i , ~
F_2(i)  =\frac{1}{2} \langle n_z(i)\sin (\omega_N t )\rangle_i 
\end{equation}
\begin{equation}
F_3(i)  =  \langle n^2_z(i)\sin^2 (\frac{\omega_N t}{2})\rangle_i , ~
F_4(i)  =  1 - F_1(i) - F_3(i)
\end{equation}
where $i$ refers to site one or two and angular brackets refer to averaging over total field distribution $W$.
There are three unique averages that need to be calculated:
\begin{eqnarray}
I_1(i)  &=& \langle \sin^2 (\frac{\omega_N  t}{2})\rangle_i, ~~
I_2(i)  =\frac{1}{2} \langle y_N \sin (\omega_N  t )\rangle_i, {}\nonumber\\
I_3(i)  &=&{}  \langle y_N^2 \sin^2 (\frac{\omega_N  t}{2})\rangle_i,
\end{eqnarray}
where we have made the change of variable $y_N = \cos\theta_N$.
Eq. (\ref{eq:salikhov}) can be expressed in closed form though we refrain from doing so for the sake of brevity. 
We still need to integrate over time and radius:
\begin{equation}\label{eq:radial}
\textrm{MR}  = 
\frac{2}{3} \frac{1}{y_{c_1}^2} \eta \int_0^{\infty} [\rho_{S}(B)  - \rho_{S}(0) ] \int_0^{y_{c_1}}  y^2  \frac{1}{\tau_h}  e^{-t/\tau_h} \operatorname dy \operatorname dt,
\end{equation}
which is never a negative value if hyperfine coupling widths are identical.
In general results are achieved by numerical integrals over $y$ and $t$. 
We confirm that our general calculation agrees with the analytic results of Section \ref{section:fastHopping} (solid symbols in Figure \ref{fig:fastHopping2}).
 However we find that performing both integrals numerically is most practical in the intermediate to fast hopping cases.
We find that in the slow hopping regime, making a change of variable $u = \exp(- 2 y)$ improves the ease of numerical evaluation.

\subsection{Saturated Magnetoresistance}\label{section:satMR}

The singlet probabilities simplify at zero field and infinite fields.
Hence it is instructive to examine the saturated MR.
The singlet probabilities reduce to the following:
\begin{equation}
\rho_S(B\rightarrow \infty) = \frac{1}{2} (1+e^{-a_{eff}^2 t^2/4}),
\end{equation}
\begin{equation}
\rho_S(B = 0) = \frac{1}{4} +\frac{1}{12}\big[1 + 2 (1 - \frac{a_{eff}^2 t^2}{4}) e^{-a_{eff}^2 t^2/8}    \big]^2.
\end{equation}
Refs. \onlinecite{Schulten1978}, \onlinecite{Salikhov1984}, and \onlinecite{Rodgers2007} provide generalizations for when the two sites are of different hyperfine species.
For slow hopping we find that it is favorable to perform the time integral of Eq. (\ref{eq:radial}) \emph{first} which can be done analytically though we omit it here because the expression is cumbersome. The integral over $u$ is then conducted numerically.
Nevertheless the extreme hopping MR converges to the following expressions:
\begin{equation}
\textrm{MR}_{sat}(v_0 \rightarrow 0)  = 
\frac{1}{27} y_{c_1} \eta,
\end{equation}
\begin{equation}
\textrm{MR}_{sat}(v_0 \rightarrow \infty)  = 
0,
\end{equation}
which qualitatively agrees with the simpler model of Ref. \onlinecite{Harmon2011}.
\begin{figure}[ptbh]
 \begin{centering}
        \includegraphics[scale = 0.75,trim = 0 0 0 0, angle = -0,clip]{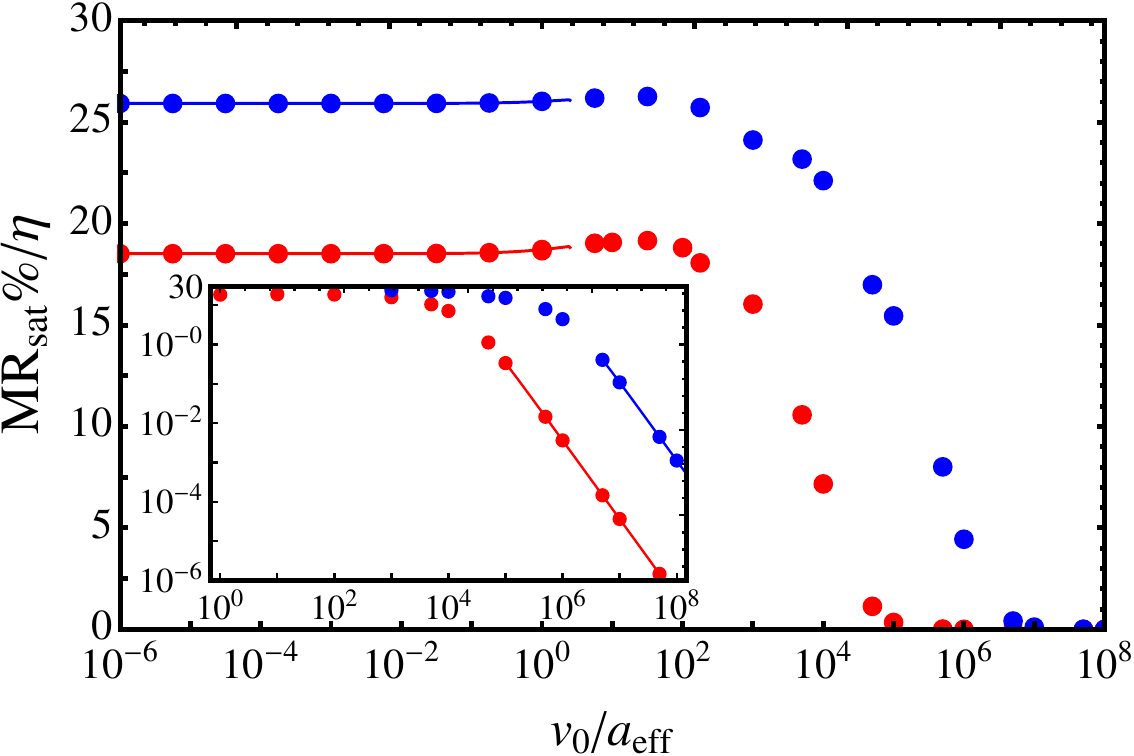}
        \caption[]
{Saturated magnetoresistance as a function of hopping rate. The inset is same as main but focuses on fast hopping rates by plotting on log-log graph. Red: $y_c = 5$; blue: $y_c = 7$. Solid lines are Taylor expansions around zero and infinite hopping rate with 50 terms. Note that the slow hopping regime extends considerably past $v_0/a_{eff} = 1$.}\label{fig:hoppingRate}
        \end{centering}
\end{figure}
Figures \ref{fig:hoppingRate} depicts the saturated MR versus the hopping rate. 
The overall shape is similar to Ref. \onlinecite{Schellekens2011}, though the decrease occurs at larger $v_0$ here (to be discussed below). 
Also in contrast, the shape is slightly non-monotonic before the sharp decline.
The hopping rate dependence highlights three interesting features:

\textit{Correspondence between critical radius and branching ratio.} The critical radius $y_c$ acts analogously to the two-site model's\cite{Bobbert2007, Schellekens2011} branching ratio $b = r_{\alpha, \beta}/r_{\alpha, e}$ where $r_{\alpha, \beta}$ is the rate from occupied site $\alpha$ to occupied site $\beta$ and $r_{\alpha, e}$ is rate from occupied site $\alpha$ to the environment (in essence, avoiding the occupied site $\beta$).
A high branching ratio entails that the polaron spin's only way to move off of $\alpha$ is to hop to $\beta$ (and can only do that if they are a singlet pair). Our critical radius acts similarly; large $y_c$ entails small site density and large inter-site spacings. 
Since sites are so far apart and the hopping rate decreases exponentially with distance, if the nearest site happens to be spin-blocked, the polaron at $\alpha$ will likely wait until the spin configuration is favorable instead of the extremely difficult further hop to a next nearest neighbor (analogous to the environment of the two-site model).
Hence hops to $\beta$ occur more frequently than hops \emph{not} to $\beta$,  much like is phenomenologically modeled by the branching ratio parameter in the two-site model.
In cases of large $y_c$ (or $b$), the saturated MR is larger, as the carrier spin must depend solely on the T-S transition (and hence the applied field) --- there is no possibility to avoid the occupied site.

\textit{Transition from fast to slow hopping occurs at an unexpected hopping rate.}  The terms `fast' and `slow' hopping are not as simple as to define as $v_0/a_{eff} \gg 1$ and $v_0/a_{eff} \ll 1$, respectively.
This is because the spatial dependence is also important and effectively decreases the hopping rate. Also $v_0 \exp(-2 y_c)/a_{eff} = 1$ is not a good measure of the criterion because most hops occur across distances less than $y_c$.
Therefore we predict that the ``slow hopping" regime is in fact applicable at faster hopping rates ($v_0$) than previously expected.
This prediction is consistent with experimental observations that large OMAR occurs with hopping rates expected to be faster than the hyperfine frequency.\cite{privateBobbert}

\textit{Limiting cases of the saturated MR for slow hopping.}  When hopping is slow, the saturated MR is \emph{independent} of \emph{both} the hyperfine coupling and the hopping rate.
This result is sensible since after long waiting times, the PP spins have sufficient time to fully mix.

\subsection{Slow Hopping Magnetoresistance Curves}\label{section:slowHoppingMR}

\begin{figure}[ptbh]
 \begin{centering}
        \includegraphics[scale = 0.8,trim = 0 0 0 0, angle = -0,clip]{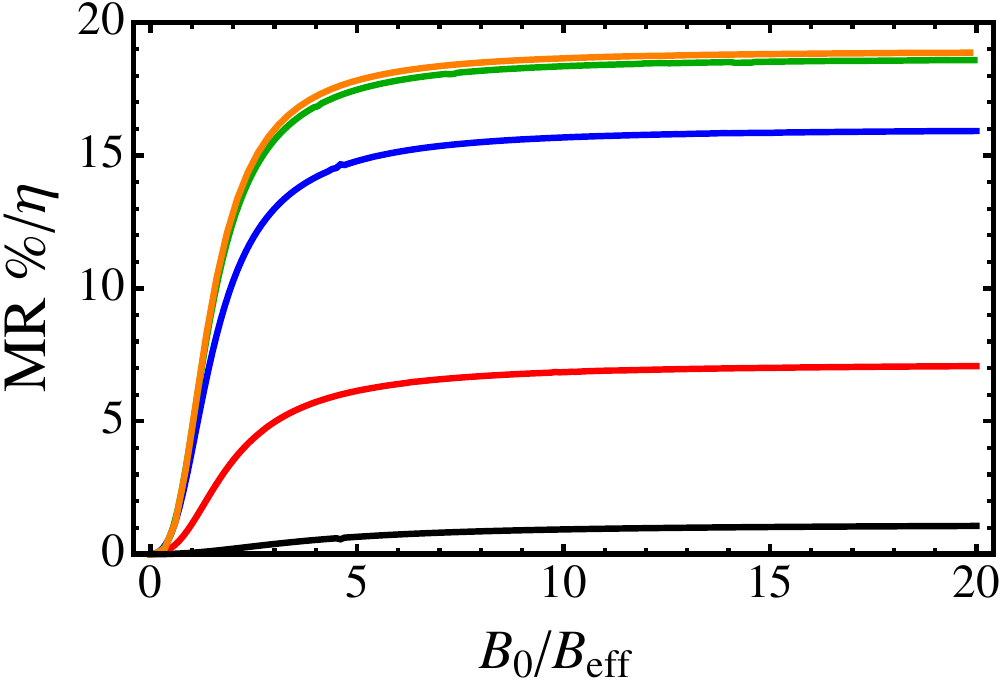}
        \caption[]
{Magnetoresistance at $y_c = 5$. Black: $v_0/a_{eff} = 5 \times 10^4$; red: $v_0/a_{eff} = 1 \times 10^4$; blue: $v_0/a_{eff} = 1 \times 10^3$; green: $v_0/a_{eff} = 1 \times 10^2$; orange: $v_0/a_{eff} = 1 \times 10^1$. The magnetoresistance is an even function of $B_0/B_{eff}$.}\label{fig:slowHoppingFig}
        \end{centering}
\end{figure}

\begin{figure}[ptbh]
 \begin{centering}
        \includegraphics[scale = 0.8,trim = 0 0 0 0, angle = -0,clip]{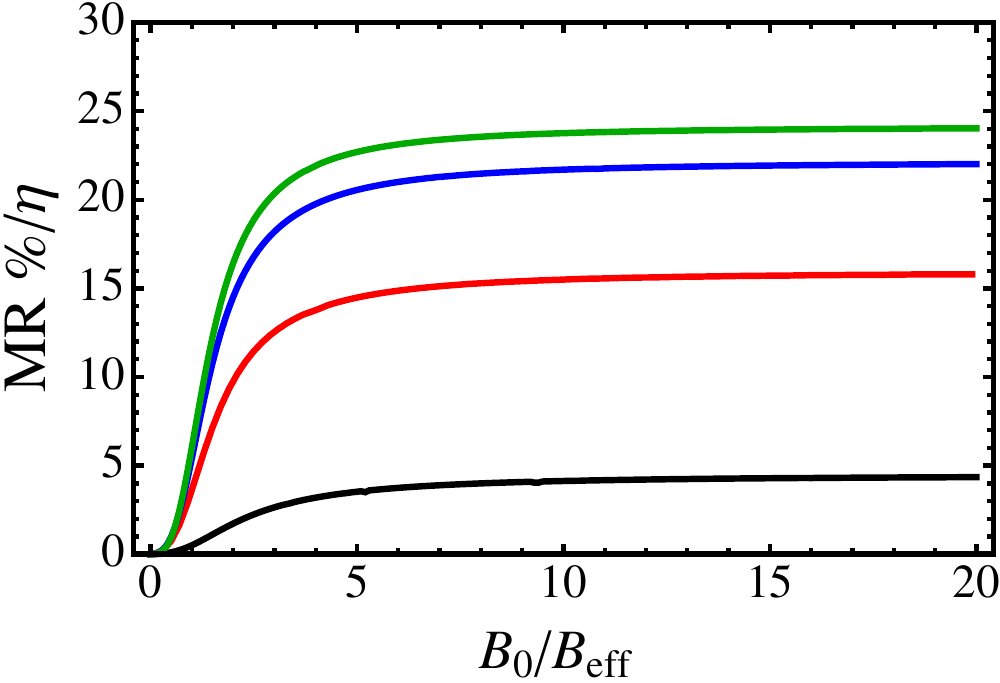}
        \caption[]
{Magnetoresistance at $y_c = 7$. Black: $v_0/a_{eff} = 1 \times 10^6$; red: $v_0/a_{eff} = 1 \times 10^5$; blue: $v_0/a_{eff} = 1 \times 10^4$; green: $v_0/a_{eff} = 1 \times 10^3$. The magnetoresistance is an even function of $B_0/B_{eff}$. }\label{fig:slowHopping7Fig}
        \end{centering}
\end{figure}

The magnetoresistance lineshapes are not fundamentally different from the discussion in Ref. \onlinecite{Harmon2011}.
Figures \ref{fig:slowHoppingFig} and \ref{fig:slowHopping7Fig} show MR traces calculated for several hopping rates at the threshold radii $y_c = 5$ and $y_c = 7$.

In the slow hopping regime (as depicted in Figure \ref{fig:hoppingRate}), the MR width is independent of hopping rate in sharp contrast to the fast hopping case (see Figure \ref{fig:fastHopping2}); the width varies linearly with the hyperfine coupling strength which also is a different behavior than seen in the fast hopping regime.
The large MR widths ($\sim 40$ mT $\gg B_{eff}$) measured by some researchers\cite{Bloom2007, Gomez2010} suggest that those scenarios were fast hopping where the hopping rate's role in MR width is indeed present.

MR is always positive, in disagreement with the ultra-small field effect observed in the simulations of Refs. \onlinecite{Kersten2011} and \onlinecite{Schellekens2011}.
At this time the source of the discrepancy between our results and their simulations is not known.
As discussed throughout this article, on other points the two approaches are in qualitative agreement.
Positive MR (ignoring ultra-small field effect) has been observed in unipolar diodes.\cite{Nguyen2010b}
It is noteworthy that spin-spin interactions also cause an ultra-small field effect to occur.\cite{Nguyen2010b, Schellekens2011, Harmon2012}
Additionally if nuclear spin moments are considered quantum mechanically, an ultra-small-field effect is expected as shown by Ref. \onlinecite{Nguyen2010b} for a single nucleus. We do not expect a small number of nuclei per site in the organic systems considered here so the semiclassical approximation is valid.\cite{Rodgers2007}

Our result suggests that organic materials with large $y_c$ (small localization length or small site concentration) yield the largest MR. 
Though increasing bias voltage tends to increase the localization length,\cite{Shklovskii1973} and therefore decrease MR in our theory, experimental observations\cite{Wagemans2010} of the bias dependence are unclear since majority and minority current injection rates possess a bias dependence for a bipolar organic device.

\section{Applicability to bipolar systems}\label{section:bipolar}

Our theory predicts solely positive MR.
This is simple to understand since applying a field suppresses hyperfine induced $T\rightarrow S$ transitions which causes a carrier to either wait for the transition or make a slower hop. This slowing down of carrier hopping leads to the increase in resistance.

However the majority of experiments have observed negative MR.\cite{Hu2007,Wagemans2010}
This is due to the presence of two types of carriers (bipolar system).
While the details of our theory do not apply in bipolar systems, we can still see qualitatively why negative MR might be dominant in a simplified model of a bipolar organic device.

Two oppositely charged polarons (an exciton) at the \emph{same} site do not contribute to the current since they will either recombine (luminesce) if a singlet or remain as an exciton if a triplet due to the large attractive Coulomb interaction (exciton disassociation is ignored for simplicity).
This is very different than the unipolar case we considered where easy formation of bipolarons encourages current flow.
If exciton formation varies between singlet and triplet e-h pairs, a similar spin-blocking mechanism emerges;
in dramatic contrast to the unipolar scenario, this time more spin mixing leads to more exciton formation which inhibits current.
An applied field suppresses spin mixing (again only considering the hyperfine interaction) which leads to less exciton formation, more current, and therefore \emph{negative} MR.
Developing a quantitative theoretical framework for bipolar OMAR based on percolation theory is a challenge for future investigations.

\section{Conclusion}\label{section:conclusion}

Testing our theory quantitatively is most tractable for high temperatures and a low density of molecular sites.
Controlling the density of sites is predicted to change the hopping rate and the resulting MR. 
Such a manipulation of site densities has been successfully employed in the past in TNF films in which conduction via $r$-percolation was measured through time-of-flight experiments.\cite{Gill1974, Rubel2004}
In these experiments the molecular density of TNF was carefully controlled by dispersing TNF in an inert polyester host.

The theory presented here has implications for MR effects in amorphous semiconductors,\cite{Movaghar1977} colloidal quantum dots,\cite{GuyotSionnest2007} spin diffusion in organic full spin valves,\cite{Bobbert2009} and MR effects in organic semi-spin valves where fringe fields from a magnetic film create a unique MR curve.\cite{Wang2011}

\section{Acknowledgements}\label{section:acknowledgements}

This work was supported by an ARO MURI. We acknowledge stimulating discussions with M. Wohlgenannt, P. A. Bobbert, and B. Koopmans.

\appendix
\section{Spin operators}\label{app:spinOperators}

We write all matrices in the singlet/triplet basis.
The spin ladder operators are
\[
S_1^+ = \frac{1}{\sqrt{2}}
\left( {\begin{array}{ccccc}
 & |S\rangle & |T_0 \rangle  & |T_+ \rangle & |T_- \rangle \\
 \langle S| & 0  & 0 &0 & 1 \\
 \langle T_0| & 0 &0 & 0 & 1\\
 \langle T_+| & -1 & 1 & 0 & 0 \\
  \langle T_-| & 0 & 0 & 0 & 0 \\
 \end{array} } \right), 
 \]
 \[
 S_2^+ = \frac{1}{\sqrt{2}}
\left( {\begin{array}{cccc}
 0  & 0 &0 & -1 \\
 0 &0 & 0 & 1 \\
1 & 1 & 0 & 0 \\
0 & 0 & 0 & 0 \\
 \end{array} } \right),
\]
and $S_i^- = S_i^{+\dagger}$. The other spin operators are $S_i^{x(y)} = \frac{S_i^{+} +(-) S_i^{-}}{2(2i)}$ and
\[
S_1^z = \frac{1}{2}
\left( {\begin{array}{cccc}
 0 & 1  & 0 & 0 \\
 1 & 0  & 0 & 0 \\
 0 & 0  & 1  & 0 \\
 0 & 0  & 0  & -1 \\
 \end{array} } \right), ~
S_2^z = \frac{1}{2}
\left( {\begin{array}{cccc}
 0 & -1  & 0 & 0 \\
 -1 & 0  & 0 & 0 \\
 0 & 0  & 1  & 0 \\
 0 & 0  & 0  & -1 \\
 \end{array} } \right).
 \]
The singlet projection operator is
 \[
P_S = 
\left( {\begin{array}{cccc}
 1 & 0  & 0 & 0 \\
 0 & 0  & 0 & 0 \\
 0 & 0  & 0  & 0 \\
 0 & 0  & 0  & 0 \\
 \end{array} } \right).
\]

\section{The $I_1$, $I_2$, and $I_3$ integrals}\label{app:integrals}

There are three unique integrals to calculate.
First,
\begin{eqnarray}
I_1 &=& 
 \frac{1}{4\pi^{1/2} a_{N}^3} e^{-\frac{1}{4}\frac{\omega_0^2}{a_{N}^2}} \times{}\\
 &&{}\int_0^{\infty} \omega_N^2 d\omega_N e^{-\frac{1}{4}\frac{ \omega_N^2 }{a_{N}^2}} \sin^2 (\frac{\omega_N  t}{2})\int_{-1}^{1} dy_N e^{\frac{1}{2}\frac{\omega_0 \omega_N y_N}{a_{N}^2}}\nonumber
\end{eqnarray}
or in dimensionless units $x_N = \omega_N/a_{N}$
\begin{eqnarray}
I_1 &=& 
 \frac{1}{4\pi^{1/2}} e^{-\frac{1}{4}\omega_0^2/a_{N}^2} \times{}\\
 &&{}\int_0^{\infty}x_N^2 dx_N e^{-\frac{1}{4}x_N^2} \sin^2 (\frac{x_N  a_{N}t}{2})\int_{-1}^{1} dy_N e^{\frac{1}{2}x_N y_N \omega_0/a_{N}}.\nonumber
\end{eqnarray}
which is
\begin{equation}
I_1 = 
\frac{1}{2} + \big[\frac{1}{2}\cos (h \tau) - \frac{\tau}{8h}   \sin (h \tau) \big]e^{-\tau^2/8},
\end{equation}
where $h =\omega_0/a_{eff}$ and $\tau = a_{eff} t$.
Also
\begin{eqnarray}
I_2 &=& 
 \frac{1}{8 \pi^{1/2}} e^{-\frac{1}{4}\omega_0^2/a_{N}^2} \times{}\\
 &&{}\int_0^{\infty}x_N^2 dx_N e^{-\frac{1}{4}x_N^2} \sin (x_N  a_{N}t )\int_{-1}^{1} dy_N y_N e^{\frac{1}{2}x_N y_N \omega_0/a_{N}}\nonumber
\end{eqnarray}
which yields
\begin{equation}
I_2 = 
\Big[ \frac{1}{2} \sin (h \tau) - \frac{1}{8 h^2} \sin (h \tau) +  \frac{\tau}{8 h} \cos (h \tau) \Big]e^{-\tau^2/8}.
\end{equation}
The last integral is
\begin{eqnarray}
I_3 &=& 
 \frac{1}{4 \pi^{1/2}} e^{-\frac{1}{4}\omega_0^2/a_{N}^2} \times{}\\
 &&{}\int_{-1}^{1} dy_N y_N^2  \int_0^{\infty}x_N^2 dx_N e^{-\frac{1}{4}x_N^2} e^{\frac{1}{2}x_N y_N \omega_0/a_{N}}\sin^2 (\frac{x_N  a_{N}t}{2} )\nonumber
\end{eqnarray}
with the result
\begin{widetext}
\begin{equation}
I_3 = I_1  -
\Big[
\frac{1 }{4 h^2} - \frac{1 }{4 h^2} \cos(h \tau) e^{- \tau^2/8} - \frac{1}{4 \sqrt{2} h^3} \textrm{D}(\sqrt{2}h) +
\frac{i}{16 \sqrt{2} h^3} e^{-2 h^2} \sqrt{\pi} (\textrm{Erf}(\frac{\tau}{2 \sqrt{2}} - i \sqrt{2} h) - \textrm{Erf}(\frac{\tau}{2 \sqrt{2}} + i \sqrt{2} h))
\Big],
\end{equation}
\end{widetext}
where $\textrm{D}(z) = e^{-z^2}\int_0^{z} e^{x^2} dx$ is Dawson's integral and $\textrm{Erf}(z) = \frac{2}{\sqrt{\pi}} \int_0^{z} e^{-x^2} dx$ is the error function.


\begin{thebibliography}{10}%
\makeatletter
\providecommand \@ifxundefined [1]{%
 \ifx #1\undefined \expandafter \@firstoftwo
 \else \expandafter \@secondoftwo
\fi
}%
\providecommand \@ifnum [1]{%
 \ifnum #1\expandafter \@firstoftwo
 \else \expandafter \@secondoftwo
\fi
}%
\providecommand \enquote [1]{``#1''}%
\providecommand \bibnamefont  [1]{#1}%
\providecommand \bibfnamefont [1]{#1}%
\providecommand \citenamefont [1]{#1}%
\providecommand\href[0]{\@sanitize\@href}%
\providecommand\@href[1]{\endgroup\@@startlink{#1}\endgroup\@@href}%
\providecommand\@@href[1]{#1\@@endlink}%
\providecommand \@sanitize [0]{\begingroup\catcode`\&12\catcode`\#12\relax}%
\@ifxundefined \pdfoutput {\@firstoftwo}{%
 \@ifnum{\z@=\pdfoutput}{\@firstoftwo}{\@secondoftwo}%
}{%
 \providecommand\@@startlink[1]{\leavevmode}%
 \providecommand\@@endlink[0]{}%
}{%
 \providecommand\@@startlink[1]{%
  \leavevmode
  \pdfstartlink
   attr{/Border[0 0 1 ]/H/I/C[0 1 1]}%
   user{/Subtype/Link/A<</Type/Action/S/URI/URI(#1)>>}%
  \relax
 }%
 \providecommand\@@endlink[0]{\pdfendlink}%
}%
\providecommand \url  [0]{\begingroup\@sanitize \@url }%
\providecommand \@url [1]{\endgroup\@href {#1}{\urlprefix}}%
\providecommand \urlprefix [0]{URL }%
\providecommand \Eprint[0]{\href }%
\@ifxundefined \urlstyle {%
  \providecommand \doi [1]{doi:\discretionary{}{}{}#1}%
}{%
  \providecommand \doi [0]{doi:\discretionary{}{}{}\begingroup
  \urlstyle{rm}\Url }%
}%
\providecommand \doibase [0]{http://dx.doi.org/}%
\providecommand \Doi[1]{\href{\doibase#1}}%
\providecommand \bibAnnote [3]{%
  \BibitemShut{#1}%
  \begin{quotation}\noindent
    \textsc{Key:}\ #2\\\textsc{Annotation:}\ #3%
  \end{quotation}%
}%
\providecommand \bibAnnoteFile [2]{%
  \IfFileExists{#2}{\bibAnnote {#1} {#2} {\input{#2}}}{}%
}%
\providecommand \typeout [0]{\immediate \write \m@ne }%
\providecommand \selectlanguage [0]{\@gobble}%
\providecommand \bibinfo [0]{\@secondoftwo}%
\providecommand \bibfield [0]{\@secondoftwo}%
\providecommand \translation [1]{[#1]}%
\providecommand \BibitemOpen[0]{}%
\providecommand \bibitemStop [0]{}%
\providecommand \bibitemNoStop [0]{.\EOS\space}%
\providecommand \EOS [0]{\spacefactor3000\relax}%
\providecommand \BibitemShut [1]{\csname bibitem#1\endcsname}%
\bibitem{Ziese2001}%
  \BibitemOpen
  \emph{\bibinfo {title} {Spin electronics}},\ edited by\ \bibinfo {editor}
  {\bibfnamefont{M.}~\bibnamefont{Ziese}}\ and\ \bibinfo {editor}
  {\bibfnamefont{M.~J.}\ \bibnamefont{Thornton}},\ \bibinfo {series} {Lecture
  Notes in Physics}, Vol.\ \bibinfo {volume} {569}\ (\bibinfo {publisher}
  {Springer-Verlag},\ \bibinfo {address} {Heidelberg},\ \bibinfo {year}
  {2001})%
  \bibAnnoteFile{NoStop}{Ziese2001}%
\bibitem{Awschalom2002}%
  \BibitemOpen
  \emph{\bibinfo {title} {Semiconductor Spintronics and Quantum Computation}},\
  edited by\ \bibinfo {editor} {\bibfnamefont{D.~D.}\ \bibnamefont{Awschalom}},
  \bibinfo {editor} {\bibfnamefont{N.}~\bibnamefont{Samarth}},\ and\ \bibinfo
  {editor} {\bibfnamefont{D.}~\bibnamefont{Loss}}\ (\bibinfo {publisher}
  {Springer Verlag},\ \bibinfo {address} {Heidelberg},\ \bibinfo {year}
  {2002})%
  \bibAnnoteFile{NoStop}{Awschalom2002}%
\bibitem{Awschalom2002SA}%
  \BibitemOpen
  \bibfield{author}{%
  \bibinfo {author} {\bibfnamefont{D.~D.}\ \bibnamefont{Awschalom}}, \bibinfo
  {author} {\bibfnamefont{M.~E.}\ \bibnamefont{Flatt\'e}},\ and\ \bibinfo
  {author} {\bibfnamefont{N.}~\bibnamefont{Samarth}},\ }%
  \bibfield{journal}{%
  \bibinfo {journal} {Scientific American}\ }%
  \textbf{\bibinfo {volume} {286}},\ \bibinfo {pages} {69} (\bibinfo {year}
  {2002})%
  \bibAnnoteFile{NoStop}{Awschalom2002SA}%
\bibitem{Naber2007}%
  \BibitemOpen
  \bibfield{author}{%
  \bibinfo {author} {\bibfnamefont{W.~J.~M.}\ \bibnamefont{Naber}}, \bibinfo
  {author} {\bibfnamefont{S.}~\bibnamefont{Faez}},\ and\ \bibinfo {author}
  {\bibfnamefont{W.~G.}\ \bibnamefont{van~der Wiel}},\ }%
  \bibfield{journal}{%
  \bibinfo {journal} {J. Phys. D: Appl. Phys.}\ }%
  \textbf{\bibinfo {volume} {40}},\ \bibinfo {pages} {R205} (\bibinfo {year}
  {2007})%
  \bibAnnoteFile{NoStop}{Naber2007}%
\bibitem{Vardeny2010}%
  \BibitemOpen
  \emph{\bibinfo {title} {Organic Spintronics}},\ edited by\ \bibinfo {editor}
  {\bibfnamefont{Z.~V.}\ \bibnamefont{Vardeny}}\ (\bibinfo {publisher} {CRC
  Press},\ \bibinfo {address} {Heidelberg},\ \bibinfo {year} {2010})%
  \bibAnnoteFile{NoStop}{Vardeny2010}%
\bibitem{Bergenti2011}%
  \BibitemOpen
  \bibfield{author}{%
  \bibinfo {author} {\bibfnamefont{I.}~\bibnamefont{Bergenti}}, \bibinfo
  {author} {\bibfnamefont{V.}~\bibnamefont{Dediu}}, \bibinfo {author}
  {\bibfnamefont{M.}~\bibnamefont{Prezioso}},\ and\ \bibinfo {author}
  {\bibfnamefont{A.}~\bibnamefont{Riminucci}},\ }%
  \bibfield{journal}{%
  \bibinfo {journal} {Phil. Trans. R. Soc. A}\ }%
  \textbf{\bibinfo {volume} {369}},\ \bibinfo {pages} {3054} (\bibinfo {year}
  {2011})%
  \bibAnnoteFile{NoStop}{Bergenti2011}%
\bibitem{Hauff2006}%
  \BibitemOpen
  \bibfield{author}{%
  \bibinfo {author} {\bibfnamefont{E.}~\bibnamefont{von Hauff}}, \bibinfo
  {author} {\bibfnamefont{C.}~\bibnamefont{Deibel}},\ and\ \bibinfo {author}
  {\bibfnamefont{V.}~\bibnamefont{Dyakonov}},\ }%
  in\ \emph{\bibinfo {booktitle} {Charge Transport in Disordered Solids}},\
  \bibinfo {editor} {edited by\ \bibinfo {editor}
  {\bibfnamefont{S.}~\bibnamefont{Baranovski}}}\ (\bibinfo {publisher} {John
  Wiley and Sons},\ \bibinfo {year} {2006})\ pp.\ \bibinfo {pages} {267--305}%
  \bibAnnoteFile{NoStop}{Hauff2006}%
\bibitem{Flatte2000b}%
  \BibitemOpen
  \bibfield{author}{%
  \bibinfo {author} {\bibfnamefont{M.~E.}\ \bibnamefont{Flatt\'e}}\ and\
  \bibinfo {author} {\bibfnamefont{J.~M.}\ \bibnamefont{Byers}},\ }%
  \bibfield{journal}{%
  \Doi{10.1103/PhysRevLett.84.4220}{\bibinfo {journal} {Phys. Rev. Lett.}}\ }%
  \textbf{\bibinfo {volume} {84}},\ \bibinfo {pages} {4220} (\bibinfo {month}
  {May}\ \bibinfo {year} {2000})%
  \bibAnnoteFile{NoStop}{Flatte2000b}%
\bibitem{Awschalom2007}%
  \BibitemOpen
  \bibfield{author}{%
  \bibinfo {author} {\bibfnamefont{D.~D.}\ \bibnamefont{Awschalom}}\ and\
  \bibinfo {author} {\bibfnamefont{M.~E.}\ \bibnamefont{Flatt\'e}},\ }%
  \bibfield{journal}{%
  \bibinfo {journal} {Nature Physics}\ }%
  \textbf{\bibinfo {volume} {3}},\ \bibinfo {pages} {153} (\bibinfo {year}
  {2007})%
  \bibAnnoteFile{NoStop}{Awschalom2007}%
\bibitem{Kalinowski2003}%
  \BibitemOpen
  \bibfield{author}{%
  \bibinfo {author} {\bibfnamefont{J.}~\bibnamefont{Kalinowski}}, \bibinfo
  {author} {\bibfnamefont{M.}~\bibnamefont{Cocchi}}, \bibinfo {author}
  {\bibfnamefont{D.}~\bibnamefont{Virgili}}, \bibinfo {author}
  {\bibfnamefont{P.~D.}\ \bibnamefont{Marco}},\ and\ \bibinfo {author}
  {\bibfnamefont{V.}~\bibnamefont{Fattori}},\ }%
  \bibfield{journal}{%
  \bibinfo {journal} {Chem. Phys. Lett.}\ }%
  \textbf{\bibinfo {volume} {380}},\ \bibinfo {pages} {710} (\bibinfo {year}
  {2003})%
  \bibAnnoteFile{NoStop}{Kalinowski2003}%
\bibitem{Francis2004}%
  \BibitemOpen
  \bibfield{author}{%
  \bibinfo {author} {\bibfnamefont{T.~L.}\ \bibnamefont{Francis}}, \bibinfo
  {author} {\bibfnamefont{O.}~\bibnamefont{Mermer}}, \bibinfo {author}
  {\bibfnamefont{G.}~\bibnamefont{Veeraraghavan}},\ and\ \bibinfo {author}
  {\bibfnamefont{M.}~\bibnamefont{Wohlgenannt}},\ }%
  \bibfield{journal}{%
  \bibinfo {journal} {New Journal of Physics}\ }%
  \textbf{\bibinfo {volume} {6}},\ \bibinfo {pages} {185} (\bibinfo {year}
  {2004})%
  \bibAnnoteFile{NoStop}{Francis2004}%
\bibitem{Prigodin2006}%
  \BibitemOpen
  \bibfield{author}{%
  \bibinfo {author} {\bibfnamefont{V.~N.}\ \bibnamefont{Prigodin}}, \bibinfo
  {author} {\bibfnamefont{J.~D.}\ \bibnamefont{Bergeson}}, \bibinfo {author}
  {\bibfnamefont{D.~M.}\ \bibnamefont{Lincoln}},\ and\ \bibinfo {author}
  {\bibfnamefont{A.~J.}\ \bibnamefont{Epstein}},\ }%
  \bibfield{journal}{%
  \bibinfo {journal} {Synthetic Metals}\ }%
  \textbf{\bibinfo {volume} {156}},\ \bibinfo {pages} {757} (\bibinfo {year}
  {2006})%
  \bibAnnoteFile{NoStop}{Prigodin2006}%
\bibitem{Desai2007}%
  \BibitemOpen
  \bibfield{author}{%
  \bibinfo {author} {\bibfnamefont{P.}~\bibnamefont{Desai}}, \bibinfo {author}
  {\bibfnamefont{P.}~\bibnamefont{Shakya}}, \bibinfo {author}
  {\bibfnamefont{T.}~\bibnamefont{Kreouzis}},\ and\ \bibinfo {author}
  {\bibfnamefont{W.~P.}\ \bibnamefont{Gillin}},\ }%
  \bibfield{journal}{%
  \bibinfo {journal} {Phys. Rev. B}\ }%
  \textbf{\bibinfo {volume} {76}},\ \bibinfo {pages} {235202} (\bibinfo {year}
  {2007})%
  \bibAnnoteFile{NoStop}{Desai2007}%
\bibitem{Hu2007}%
  \BibitemOpen
  \bibfield{author}{%
  \bibinfo {author} {\bibfnamefont{B.}~\bibnamefont{Hu}}\ and\ \bibinfo
  {author} {\bibfnamefont{Y.}~\bibnamefont{Wu}},\ }%
  \bibfield{journal}{%
  \bibinfo {journal} {Nature Materials}\ }%
  \textbf{\bibinfo {volume} {6}},\ \bibinfo {pages} {985} (\bibinfo {year}
  {2007})%
  \bibAnnoteFile{NoStop}{Hu2007}%
\bibitem{Bloom2007}%
  \BibitemOpen
  \bibfield{author}{%
  \bibinfo {author} {\bibfnamefont{F.~L.}\ \bibnamefont{Bloom}}, \bibinfo
  {author} {\bibfnamefont{W.}~\bibnamefont{Wagemans}}, \bibinfo {author}
  {\bibfnamefont{M.}~\bibnamefont{Kemerink}},\ and\ \bibinfo {author}
  {\bibfnamefont{B.}~\bibnamefont{Koopmans}},\ }%
  \bibfield{journal}{%
  \bibinfo {journal} {Phys. Rev. Lett.}\ }%
  \textbf{\bibinfo {volume} {99}},\ \bibinfo {pages} {257201} (\bibinfo {year}
  {2007})%
  \bibAnnoteFile{NoStop}{Bloom2007}%
\bibitem{Bobbert2007}%
  \BibitemOpen
  \bibfield{author}{%
  \bibinfo {author} {\bibfnamefont{P.~A.}\ \bibnamefont{Bobbert}}, \bibinfo
  {author} {\bibfnamefont{T.~D.}\ \bibnamefont{Nguyen}}, \bibinfo {author}
  {\bibfnamefont{F.~W.~A.}\ \bibnamefont{van Oost}}, \bibinfo {author}
  {\bibfnamefont{B.}~\bibnamefont{Koopmans}},\ and\ \bibinfo {author}
  {\bibfnamefont{M.}~\bibnamefont{Wohlgenannt}},\ }%
  \bibfield{journal}{%
  \bibinfo {journal} {Phys. Rev. Lett.}\ }%
  \textbf{\bibinfo {volume} {99}},\ \bibinfo {pages} {216801} (\bibinfo {year}
  {2007})%
  \bibAnnoteFile{NoStop}{Bobbert2007}%
\bibitem{Bergeson2008}%
  \BibitemOpen
  \bibfield{author}{%
  \bibinfo {author} {\bibfnamefont{J.~D.}\ \bibnamefont{Bergeson}}, \bibinfo
  {author} {\bibfnamefont{V.~N.}\ \bibnamefont{Prigodin}}, \bibinfo {author}
  {\bibfnamefont{D.~M.}\ \bibnamefont{Lincoln}},\ and\ \bibinfo {author}
  {\bibfnamefont{A.~J.}\ \bibnamefont{Epstein}},\ }%
  \bibfield{journal}{%
  \bibinfo {journal} {Phys. Rev. Lett.}\ }%
  \textbf{\bibinfo {volume} {100}},\ \bibinfo {pages} {067201} (\bibinfo {year}
  {2008})%
  \bibAnnoteFile{NoStop}{Bergeson2008}%
\bibitem{Wagemans2010}%
  \BibitemOpen
  \bibfield{author}{%
  \bibinfo {author} {\bibfnamefont{W.}~\bibnamefont{Wagemans}}\ and\ \bibinfo
  {author} {\bibfnamefont{B.}~\bibnamefont{Koopmans}},\ }%
  \bibfield{journal}{%
  \bibinfo {journal} {Phys. Satus Solidi B}\ }%
  \textbf{\bibinfo {volume} {248}},\ \bibinfo {pages} {1029} (\bibinfo {year}
  {2011})%
  \bibAnnoteFile{NoStop}{Wagemans2010}%
\bibitem{Nguyen2010}%
  \BibitemOpen
  \bibfield{author}{%
  \bibinfo {author} {\bibfnamefont{T.~D.}\ \bibnamefont{Nguyen}}, \bibinfo
  {author} {\bibfnamefont{G.}~\bibnamefont{Hukic-Markosian}}, \bibinfo {author}
  {\bibfnamefont{F.}~\bibnamefont{Wang}}, \bibinfo {author}
  {\bibfnamefont{L.}~\bibnamefont{Wojcik}}, \bibinfo {author}
  {\bibfnamefont{X.-G.}\ \bibnamefont{Li}}, \bibinfo {author}
  {\bibfnamefont{E.}~\bibnamefont{Ehrenfreund}},\ and\ \bibinfo {author}
  {\bibfnamefont{Z.~V.}\ \bibnamefont{Vardeny}},\ }%
  \bibfield{journal}{%
  \bibinfo {journal} {Nature Materials}\ }%
  \textbf{\bibinfo {volume} {9}},\ \bibinfo {pages} {345} (\bibinfo {year}
  {2010})%
  \bibAnnoteFile{NoStop}{Nguyen2010}%
\bibitem{Veeraraghavan2007}%
  \BibitemOpen
  \bibfield{author}{%
  \bibinfo {author} {\bibfnamefont{G.}~\bibnamefont{Veeraraghavan}}, \bibinfo
  {author} {\bibfnamefont{T.~D.}\ \bibnamefont{Nguyen}}, \bibinfo {author}
  {\bibfnamefont{Y.}~\bibnamefont{Sheng}}, \bibinfo {author}
  {\bibfnamefont{O.}~\bibnamefont{Mermer}},\ and\ \bibinfo {author}
  {\bibfnamefont{M.}~\bibnamefont{Wohlgenannt}},\ }%
  \bibfield{journal}{%
  \bibinfo {journal} {IEEE Transactions on Electron Devices}\ }%
  \textbf{\bibinfo {volume} {54}},\ \bibinfo {pages} {1571} (\bibinfo {year}
  {2007})%
  \bibAnnoteFile{NoStop}{Veeraraghavan2007}%
\bibitem{Mermer2005a}%
  \BibitemOpen
  \bibfield{author}{%
  \bibinfo {author} {\bibfnamefont{O.}~\bibnamefont{Mermer}}, \bibinfo {author}
  {\bibfnamefont{G.}~\bibnamefont{Veeraraghavan}}, \bibinfo {author}
  {\bibfnamefont{T.~L.}\ \bibnamefont{Francis}}, \bibinfo {author}
  {\bibfnamefont{Y.}~\bibnamefont{Sheng}}, \bibinfo {author}
  {\bibfnamefont{D.~T.}\ \bibnamefont{Nguyen}}, \bibinfo {author}
  {\bibfnamefont{M.}~\bibnamefont{Wohlgenannt}}, \bibinfo {author}
  {\bibfnamefont{A.}~\bibnamefont{K\"ohler}}, \bibinfo {author}
  {\bibfnamefont{M.~K.}\ \bibnamefont{Al-Suti}},\ and\ \bibinfo {author}
  {\bibfnamefont{M.~S.}\ \bibnamefont{Khan}},\ }%
  \bibfield{journal}{%
  \bibinfo {journal} {Phys. Rev. B}\ }%
  \textbf{\bibinfo {volume} {72}},\ \bibinfo {pages} {205202} (\bibinfo {year}
  {2005})%
  \bibAnnoteFile{NoStop}{Mermer2005a}%
\bibitem{Shklovskii1984}%
  \BibitemOpen
  \bibfield{author}{%
  \bibinfo {author} {\bibfnamefont{B.~I.}\ \bibnamefont{Shklovskii}}\ and\
  \bibinfo {author} {\bibfnamefont{A.~L.}\ \bibnamefont{Efros}},\ }%
  \emph{\bibinfo {title} {Electronic Properties of Doped Semiconductors}}\
  (\bibinfo {publisher} {Springer},\ \bibinfo {address} {Heidelberg},\ \bibinfo
  {year} {1984})%
  \bibAnnoteFile{NoStop}{Shklovskii1984}%
\bibitem{Baranovski2006}%
  \BibitemOpen
  \bibfield{author}{%
  \bibinfo {author} {\bibfnamefont{S.}~\bibnamefont{Baranovski}}\ and\ \bibinfo
  {author} {\bibfnamefont{O.}~\bibnamefont{Rubel}},\ }%
  in\ \emph{\bibinfo {booktitle} {Charge Transport in Disordered Solids}},\
  \bibinfo {editor} {edited by\ \bibinfo {editor}
  {\bibfnamefont{S.}~\bibnamefont{Baranovski}}}\ (\bibinfo {publisher} {John
  Wiley and Sons},\ \bibinfo {year} {2006})\ pp.\ \bibinfo {pages} {221--266}%
  \bibAnnoteFile{NoStop}{Baranovski2006}%
\bibitem{Harmon2011}%
  \BibitemOpen
  \bibfield{author}{%
  \bibinfo {author} {\bibfnamefont{N.~J.}\ \bibnamefont{Harmon}}\ and\ \bibinfo
  {author} {\bibfnamefont{M.~E.}\ \bibnamefont{Flatt\'e}},\ }%
  \bibinfo {journal} {arXiv:1106.3040v1}%
  \bibAnnoteFile{NoStop}{Harmon2011}%
\bibitem{Wagemans2008}%
  \BibitemOpen
\bibfield{journal}{%
    }%
  \bibfield{author}{%
  \bibinfo {author} {\bibfnamefont{W.}~\bibnamefont{Wagemans}}, \bibinfo
  {author} {\bibfnamefont{F.~L.}\ \bibnamefont{Bloom}}, \bibinfo {author}
  {\bibfnamefont{P.~A.}\ \bibnamefont{Bobbert}}, \bibinfo {author}
  {\bibfnamefont{M.}~\bibnamefont{Wohlgenannt}},\ and\ \bibinfo {author}
  {\bibfnamefont{B.}~\bibnamefont{Koopmans}},\ }%
  \bibfield{journal}{%
  \bibinfo {journal} {J. Appl. Phys.}\ }%
  \textbf{\bibinfo {volume} {103}},\ \bibinfo {pages} {07F303} (\bibinfo {year}
  {2008})%
  \bibAnnoteFile{NoStop}{Wagemans2008}%
\bibitem{Arkhipov2002}%
  \BibitemOpen
  \bibfield{author}{%
  \bibinfo {author} {\bibfnamefont{V.~I.}\ \bibnamefont{Arkhipov}},\ }%
  \bibfield{journal}{%
  \bibinfo {journal} {Phys. Rev. B}\ }%
  \textbf{\bibinfo {volume} {65}},\ \bibinfo {pages} {165110} (\bibinfo {year}
  {2002})%
  \bibAnnoteFile{NoStop}{Arkhipov2002}%
\bibitem{Schellekens2011}%
  \BibitemOpen
  \bibfield{author}{%
  \bibinfo {author} {\bibfnamefont{A.~J.}\ \bibnamefont{Schellekens}}, \bibinfo
  {author} {\bibfnamefont{W.}~\bibnamefont{Wagemans}}, \bibinfo {author}
  {\bibfnamefont{S.~P.}\ \bibnamefont{Kersten}}, \bibinfo {author}
  {\bibfnamefont{P.~A.}\ \bibnamefont{Bobbert}},\ and\ \bibinfo {author}
  {\bibfnamefont{B.}~\bibnamefont{Koopmans}},\ }%
  \bibfield{journal}{%
  \bibinfo {journal} {Phys. Rev. B}\ }%
  \textbf{\bibinfo {volume} {84}},\ \bibinfo {pages} {075204} (\bibinfo {year}
  {2011})%
  \bibAnnoteFile{NoStop}{Schellekens2011}%
\bibitem{Miller1960}%
  \BibitemOpen
  \bibfield{author}{%
  \bibinfo {author} {\bibfnamefont{A.}~\bibnamefont{Miller}}\ and\ \bibinfo
  {author} {\bibfnamefont{E.}~\bibnamefont{Abrahams}},\ }%
  \bibfield{journal}{%
  \bibinfo {journal} {Phys. Rev.}\ }%
  \textbf{\bibinfo {volume} {120}},\ \bibinfo {pages} {745} (\bibinfo {year}
  {1960})%
  \bibAnnoteFile{NoStop}{Miller1960}%
\bibitem{Ambegaokar1971}%
  \BibitemOpen
  \bibfield{author}{%
  \bibinfo {author} {\bibfnamefont{V.}~\bibnamefont{Ambegaokar}}, \bibinfo
  {author} {\bibfnamefont{B.~I.}\ \bibnamefont{Halperin}},\ and\ \bibinfo
  {author} {\bibfnamefont{J.~S.}\ \bibnamefont{Langer}},\ }%
  \bibfield{journal}{%
  \bibinfo {journal} {Phys. Rev. B}\ }%
  \textbf{\bibinfo {volume} {4}},\ \bibinfo {pages} {2612} (\bibinfo {year}
  {1971})%
  \bibAnnoteFile{NoStop}{Ambegaokar1971}%
\bibitem{Pollak1972}%
  \BibitemOpen
  \bibfield{author}{%
  \bibinfo {author} {\bibfnamefont{M.}~\bibnamefont{Pollak}},\ }%
  \bibfield{journal}{%
  \bibinfo {journal} {J. Non-Crystalline Solids}\ }%
  \textbf{\bibinfo {volume} {11}},\ \bibinfo {pages} {1} (\bibinfo {year}
  {1972})%
  \bibAnnoteFile{NoStop}{Pollak1972}%
\bibitem{Gill1974}%
  \BibitemOpen
  \bibfield{author}{%
  \bibinfo {author} {\bibfnamefont{W.}~\bibnamefont{Gill}},\ }%
  \bibfield{journal}{%
  \bibinfo {journal} {Proceedings of the Fifth International Conference on
  Amorphous and Liquid Semiconductors}\ }%
  \textbf{\bibinfo {volume} {7222}},\ \bibinfo {pages} {901} (\bibinfo {year}
  {1974})%
  \bibAnnoteFile{NoStop}{Gill1974}%
\bibitem{Rubel2004}%
  \BibitemOpen
  \bibfield{author}{%
  \bibinfo {author} {\bibfnamefont{O.}~\bibnamefont{Rubel}}, \bibinfo {author}
  {\bibfnamefont{S.~D.}\ \bibnamefont{Baranovskii}}, \bibinfo {author}
  {\bibfnamefont{P.}~\bibnamefont{Thomas}},\ and\ \bibinfo {author}
  {\bibfnamefont{S.}~\bibnamefont{Yamasaki}},\ }%
  \bibfield{journal}{%
  \bibinfo {journal} {Phys. Rev. B}\ }%
  \textbf{\bibinfo {volume} {69}},\ \bibinfo {pages} {014206} (\bibinfo {year}
  {2004})%
  \bibAnnoteFile{NoStop}{Rubel2004}%
\bibitem{Bussac1993}%
  \BibitemOpen
  \bibfield{author}{%
  \bibinfo {author} {\bibfnamefont{M.~N.}\ \bibnamefont{Bussac}}\ and\ \bibinfo
  {author} {\bibfnamefont{L.}~\bibnamefont{Zuppiroli}},\ }%
  \bibfield{journal}{%
  \bibinfo {journal} {Phys. Rev. B}\ }%
  \textbf{\bibinfo {volume} {47}},\ \bibinfo {pages} {5493} (\bibinfo {year}
  {1993})%
  \bibAnnoteFile{NoStop}{Bussac1993}%
\bibitem{Osaka1979}%
  \BibitemOpen
  \bibfield{author}{%
  \bibinfo {author} {\bibfnamefont{Y.}~\bibnamefont{Osaka}},\ }%
  \bibfield{journal}{%
  \bibinfo {journal} {J. Phys. Soc. Japan}\ }%
  \textbf{\bibinfo {volume} {47}},\ \bibinfo {pages} {729} (\bibinfo {year}
  {1979})%
  \bibAnnoteFile{NoStop}{Osaka1979}%
\bibitem{Schulten1978}%
  \BibitemOpen
  \bibfield{author}{%
  \bibinfo {author} {\bibfnamefont{K.}~\bibnamefont{Schulten}}\ and\ \bibinfo
  {author} {\bibfnamefont{P.~G.}\ \bibnamefont{Wolynes}},\ }%
  \bibfield{journal}{%
  \bibinfo {journal} {J. Chem. Phys.}\ }%
  \textbf{\bibinfo {volume} {68}},\ \bibinfo {pages} {3292} (\bibinfo {year}
  {1978})%
  \bibAnnoteFile{NoStop}{Schulten1978}%
\bibitem{Wagemans2011a}%
  \BibitemOpen
  \bibfield{author}{%
  \bibinfo {author} {\bibfnamefont{W.}~\bibnamefont{Wagemans}}, \bibinfo
  {author} {\bibfnamefont{P.}~\bibnamefont{Janssen}}, \bibinfo {author}
  {\bibfnamefont{A.~J.}\ \bibnamefont{Schellekens}}, \bibinfo {author}
  {\bibfnamefont{F.~L.}\ \bibnamefont{Bloom}}, \bibinfo {author}
  {\bibfnamefont{P.~A.}\ \bibnamefont{Bobbert}},\ and\ \bibinfo {author}
  {\bibfnamefont{B.}~\bibnamefont{Koopmans}},\ }%
  \bibfield{journal}{%
  \bibinfo {journal} {SPIN}\ }%
  \textbf{\bibinfo {volume} {1}},\ \bibinfo {pages} {93} (\bibinfo {year}
  {2011})%
  \bibAnnoteFile{NoStop}{Wagemans2011a}%
\bibitem{Werner1977}%
  \BibitemOpen
  \bibfield{author}{%
  \bibinfo {author} {\bibfnamefont{H.~J.}\ \bibnamefont{Werner}}, \bibinfo
  {author} {\bibfnamefont{Z.}~\bibnamefont{Schulten}},\ and\ \bibinfo {author}
  {\bibfnamefont{K.}~\bibnamefont{Schulten}},\ }%
  \bibfield{journal}{%
  \bibinfo {journal} {The Journal of Chemical Physics}\ }%
  \textbf{\bibinfo {volume} {67}},\ \bibinfo {pages} {646} (\bibinfo {year}
  {1977})%
  \bibAnnoteFile{NoStop}{Werner1977}%
\bibitem{Harmon2012}%
  \BibitemOpen
  \bibfield{author}{%
  \bibinfo {author} {\bibfnamefont{N.~J.}\ \bibnamefont{Harmon}}\ and\ \bibinfo
  {author} {\bibfnamefont{M.~E.}\ \bibnamefont{Flatt\'e}}\ }%
  \bibinfo {note} {in preparation}%
  \bibAnnoteFile{NoStop}{Harmon2012}%
\bibitem{Flory1969}%
  \BibitemOpen
  \bibfield{author}{%
  \bibinfo {author} {\bibfnamefont{P.~J.}\ \bibnamefont{Flory}},\ }%
  \emph{\bibinfo {title} {Statistical Mechanics of Chain Molecules}}\ (\bibinfo
  {publisher} {Interscience Publishers},\ \bibinfo {year} {1969})%
  \bibAnnoteFile{NoStop}{Flory1969}%
\bibitem{Rodgers2007}%
  \BibitemOpen
  \bibfield{author}{%
  \bibinfo {author} {\bibfnamefont{C.}~\bibnamefont{Rodgers}},\ }%
  \emph{\bibinfo {title} {Magnetic Field Effects in Chemical Systems}},\ Ph.D.
  thesis,\ \bibinfo {school} {University of Oxford} (\bibinfo {year} {2007})%
  \bibAnnoteFile{NoStop}{Rodgers2007}%
\bibitem{Knapp1979}%
  \BibitemOpen
  \bibfield{author}{%
  \bibinfo {author} {\bibfnamefont{E.~W.}\ \bibnamefont{Knapp}}\ and\ \bibinfo
  {author} {\bibfnamefont{K.}~\bibnamefont{Schulten}},\ }%
  \bibfield{journal}{%
  \bibinfo {journal} {J. Chem. Phys.}\ }%
  \textbf{\bibinfo {volume} {71}},\ \bibinfo {pages} {1878} (\bibinfo {year}
  {1979})%
  \bibAnnoteFile{NoStop}{Knapp1979}%
\bibitem{Haberkorn1977}%
  \BibitemOpen
  \bibfield{author}{%
  \bibinfo {author} {\bibfnamefont{R.}~\bibnamefont{Haberkorn}},\ }%
  \bibfield{journal}{%
  \bibinfo {journal} {Chemical Physics}\ }%
  \textbf{\bibinfo {volume} {24}},\ \bibinfo {pages} {111} (\bibinfo {year}
  {1977})%
  \bibAnnoteFile{NoStop}{Haberkorn1977}%
\bibitem{Slichter1963}%
  \BibitemOpen
  \bibfield{author}{%
  \bibinfo {author} {\bibfnamefont{C.~P.}\ \bibnamefont{Slichter}},\ }%
  \emph{\bibinfo {title} {Principles of Magnetic Resonance}}\ (\bibinfo
  {publisher} {Harper and Row},\ \bibinfo {address} {New York},\ \bibinfo
  {year} {1963})%
  \bibAnnoteFile{NoStop}{Slichter1963}%
\bibitem{Salikhov1984}%
  \BibitemOpen
  \bibfield{author}{%
  \bibinfo {author} {\bibfnamefont{K.~M.}\ \bibnamefont{Salikhov}}, \bibinfo
  {author} {\bibfnamefont{Y.~N.}\ \bibnamefont{Molin}}, \bibinfo {author}
  {\bibfnamefont{R.~Z.}\ \bibnamefont{Sagdeev}},\ and\ \bibinfo {author}
  {\bibfnamefont{A.~L.}\ \bibnamefont{Buchachenko}},\ }%
  in\ \emph{\bibinfo {booktitle} {Spin Polarization and Magnetic Effects in
  Radical Reactions}},\ \bibinfo {editor} {edited by\ \bibinfo {editor}
  {\bibfnamefont{Y.~N.}\ \bibnamefont{Molin}}}\ (\bibinfo {publisher}
  {Elsevier},\ \bibinfo {year} {1984})\ Chap.\ \bibinfo {chapter} {The theory
  of radical recombination}, pp.\ \bibinfo {pages} {32--116}%
  \bibAnnoteFile{NoStop}{Salikhov1984}%
\bibitem{Timmel1998}%
  \BibitemOpen
  \bibfield{author}{%
  \bibinfo {author} {\bibfnamefont{C.~R.}\ \bibnamefont{Timmel}}, \bibinfo
  {author} {\bibfnamefont{U.}~\bibnamefont{Till}}, \bibinfo {author}
  {\bibfnamefont{B.}~\bibnamefont{Brocklehurst}}, \bibinfo {author}
  {\bibfnamefont{K.~A.}\ \bibnamefont{McLaughlin}},\ and\ \bibinfo {author}
  {\bibfnamefont{P.~J.}\ \bibnamefont{Hore}},\ }%
  \bibfield{journal}{%
  \bibinfo {journal} {Molecular Physics}\ }%
  \textbf{\bibinfo {volume} {95}},\ \bibinfo {pages} {71} (\bibinfo {year}
  {1998})%
  \bibAnnoteFile{NoStop}{Timmel1998}%
\bibitem{privateBobbert}%
  \BibitemOpen
  \ \bibinfo {note} {private communications with P. A. Bobbert}%
  \bibAnnoteFile{NoStop}{privateBobbert}%
\bibitem{Gomez2010}%
  \BibitemOpen
  \bibfield{author}{%
  \bibinfo {author} {\bibfnamefont{J.~A.}\ \bibnamefont{Gomez}}, \bibinfo
  {author} {\bibfnamefont{F.}~\bibnamefont{Nuesch}}, \bibinfo {author}
  {\bibfnamefont{L.}~\bibnamefont{Zuppiroli}},\ and\ \bibinfo {author}
  {\bibfnamefont{C.~F.~O.}\ \bibnamefont{Graeff}},\ }%
  \bibfield{journal}{%
  \bibinfo {journal} {Synthetic Metals}\ }%
  \textbf{\bibinfo {volume} {160}},\ \bibinfo {pages} {317} (\bibinfo {year}
  {2010})%
  \bibAnnoteFile{NoStop}{Gomez2010}%
\bibitem{Kersten2011}%
  \BibitemOpen
  \bibfield{author}{%
  \bibinfo {author} {\bibfnamefont{S.~P.}\ \bibnamefont{Kersten}}, \bibinfo
  {author} {\bibfnamefont{A.~J.}\ \bibnamefont{Schellekens}}, \bibinfo {author}
  {\bibfnamefont{B.}~\bibnamefont{Koopmans}},\ and\ \bibinfo {author}
  {\bibfnamefont{P.~A.}\ \bibnamefont{Bobbert}},\ }%
  \bibfield{journal}{%
  \bibinfo {journal} {Phys. Rev. Lett.}\ }%
  \textbf{\bibinfo {volume} {106}},\ \bibinfo {pages} {197402} (\bibinfo {year}
  {2011})%
  \bibAnnoteFile{NoStop}{Kersten2011}%
\bibitem{Nguyen2010b}%
  \BibitemOpen
  \bibfield{author}{%
  \bibinfo {author} {\bibfnamefont{T.~D.}\ \bibnamefont{Nguyen}}, \bibinfo
  {author} {\bibfnamefont{B.~R.}\ \bibnamefont{Gautam}}, \bibinfo {author}
  {\bibfnamefont{E.}~\bibnamefont{Ehrenfreund}},\ and\ \bibinfo {author}
  {\bibfnamefont{Z.~V.}\ \bibnamefont{Vardeny}},\ }%
  \bibfield{journal}{%
  \bibinfo {journal} {Phys. Rev. Lett.}\ }%
  \textbf{\bibinfo {volume} {105}},\ \bibinfo {pages} {166804} (\bibinfo {year}
  {2010})%
  \bibAnnoteFile{NoStop}{Nguyen2010b}%
\bibitem{Shklovskii1973}%
  \BibitemOpen
  \bibfield{author}{%
  \bibinfo {author} {\bibfnamefont{B.~I.}\ \bibnamefont{Shklovskii}},\ }%
  \bibfield{journal}{%
  \bibinfo {journal} {Sov. Phys. Semicond.}\ }%
  \textbf{\bibinfo {volume} {6}},\ \bibinfo {pages} {1964} (\bibinfo {year}
  {1973})%
  \bibAnnoteFile{NoStop}{Shklovskii1973}%
\bibitem{Movaghar1977}%
  \BibitemOpen
  \bibfield{author}{%
  \bibinfo {author} {\bibfnamefont{B.}~\bibnamefont{Movaghar}}\ and\ \bibinfo
  {author} {\bibfnamefont{L.}~\bibnamefont{Schweitzer}},\ }%
  \bibfield{journal}{%
  \bibinfo {journal} {Phys. Stat. Solidi (b)}\ }%
  \textbf{\bibinfo {volume} {80}},\ \bibinfo {pages} {491} (\bibinfo {year}
  {1977})%
  \bibAnnoteFile{NoStop}{Movaghar1977}%
\bibitem{GuyotSionnest2007}%
  \BibitemOpen
  \bibfield{author}{%
  \bibinfo {author} {\bibfnamefont{P.}~\bibnamefont{Guyot-Sionnest}}, \bibinfo
  {author} {\bibfnamefont{D.}~\bibnamefont{Yu}}, \bibinfo {author}
  {\bibfnamefont{P.}~\bibnamefont{Jiang}},\ and\ \bibinfo {author}
  {\bibfnamefont{W.}~\bibnamefont{Kang}},\ }%
  \bibfield{journal}{%
  \bibinfo {journal} {J. Chem. Phys.}\ }%
  \textbf{\bibinfo {volume} {127}},\ \bibinfo {pages} {014702} (\bibinfo {year}
  {2007})%
  \bibAnnoteFile{NoStop}{GuyotSionnest2007}%
\bibitem{Bobbert2009}%
  \BibitemOpen
  \bibfield{author}{%
  \bibinfo {author} {\bibfnamefont{P.~A.}\ \bibnamefont{Bobbert}}, \bibinfo
  {author} {\bibfnamefont{W.}~\bibnamefont{Wagemans}}, \bibinfo {author}
  {\bibfnamefont{F.~W.~A.}\ \bibnamefont{Oost}}, \bibinfo {author}
  {\bibfnamefont{B.}~\bibnamefont{Koopmans}},\ and\ \bibinfo {author}
  {\bibfnamefont{M.}~\bibnamefont{Wohlgenannt}},\ }%
  \bibfield{journal}{%
  \bibinfo {journal} {Phys. Rev. Lett.}\ }%
  \textbf{\bibinfo {volume} {102}},\ \bibinfo {pages} {156604} (\bibinfo {year}
  {2009})%
  \bibAnnoteFile{NoStop}{Bobbert2009}%
\bibitem{Wang2011}%
  \BibitemOpen
  \bibfield{author}{%
  \bibinfo {author} {\bibfnamefont{F.}~\bibnamefont{Wang}}, \bibinfo {author}
  {\bibfnamefont{F.}~\bibnamefont{Maci\'a}}, \bibinfo {author}
  {\bibfnamefont{M.}~\bibnamefont{Wohlgenannt}}, \bibinfo {author}
  {\bibfnamefont{A.~D.}\ \bibnamefont{Kent}},\ and\ \bibinfo {author}
  {\bibfnamefont{M.~E.}\ \bibnamefont{Flatt\'e}},\ }%
  \bibinfo {journal} {unpublished}%
  \bibAnnoteFile{NoStop}{Wang2011}%
\end{thebibliography}

%

\end{document}